\documentclass[aps, pre, reprint, showpacs, showkeys, onecolumn, 10pt, groupedaddress, notitlepage]{revtex4-1}

\usepackage{graphicx}
\usepackage{amsmath}
\usepackage{dsfont}
\usepackage{dcolumn}
\usepackage{bm}
\usepackage{times}
\usepackage{color}
\usepackage{braket}

\newcommand{\tn}{\tabularnewline}
\newcommand{\tss}{\textsuperscript}

\begin{document}

\bibliographystyle{unsrt}

\title{Prospects for Spin-Based Quantum Computing}
\author{Christoph Kloeffel}
\affiliation{Department of Physics, University of Basel, Klingelbergstrasse 82, CH-4056 Basel, Switzerland}
\author{Daniel Loss}
\affiliation{Department of Physics, University of Basel, Klingelbergstrasse 82, CH-4056 Basel, Switzerland}
%\date{\today}

\begin{abstract}
Experimental and theoretical progress toward quantum computation with spins in quantum dots (QDs) is reviewed, with particular focus on QDs formed in GaAs heterostructures, on nanowire-based QDs, and on self-assembled QDs. We report on a remarkable evolution of the field where decoherence -- one of the main challenges for realizing quantum computers -- no longer seems to be the stumbling block it had originally been considered. General concepts, relevant quantities, and basic requirements for spin-based quantum computing are explained; opportunities and challenges of spin-orbit interaction and nuclear spins are reviewed. We discuss recent achievements, present current theoretical proposals, and make several suggestions for further experiments.  
\end{abstract}

%\pacs{}


\maketitle

\section{Introduction}

The concept of entanglement and non-locality \cite{schroedinger:naturwiss35}, one of the most striking features of quantum mechanics, has been heavily debated since the early days of the field \cite{einstein:pr35}. By now, there is abundant experimental evidence \cite{aspect:prl82} that Nature indeed does possess nonlocal aspects, in stark contrast to our everyday-life experience.

But it was only relatively recently, when Richard Feynman \cite{feynman:ijtp82, feynman:optn85}, David Deutsch \cite{deutsch:prsla85}, and other researchers in the 1980s, envisioned the idea of exploiting the quantum degrees of freedom for a novel way of information processing. The central question at the time was whether and how it is possible to efficiently simulate any finite physical system with a man-made machine. Deutsch argued that such a simulation is not possible perfectly within the classical computational framework that had been developed for decades \cite{deutsch:prsla85}. Instead, the universal computing machine should be of a quantum nature, i.e., a ``quantum computer''.

Since then, progress in different areas of research and industry tremendously influenced the advent of quantum computing. First, the booming computer industry led to major progress in semiconductor-, nano-, and laser-technology, a prerequisite for the fabrication, addressing, and manipulation of single quantum systems needed in an experimental realization. Second, several algorithms have been developed by now, such as those by Deutsch \cite{deutsch:prsla85, deutsch:prsla92}, Grover \cite{grover:prl97}, and Shor \cite{shor:pfocs94, shor:siam97}, which clearly illustrate that quantum computers, exploiting the existence of entanglement, can solve problems much faster than classical computers. A recent review on using quantum computers for quantum simulations can be found in Ref.\ \cite{brown:entropy10}. In addition, the theories of quantum complexity and entanglement are currently being established, a process which is still far from being complete. The emerging fields of nano- and quantum information science have inspired and motivated each other in various ways, today more than ever.

Shortly after the first quantum algorithms had been developed, setups have been suggested to turn quantum computing into reality. These ideas, among others, are based on quantum dots \cite{loss:pra98, imamoglu:prl99}, cold trapped ions \cite{cirac:prl95}, cavity quantum electrodynamics \cite{imamoglu:prl99, turchette:prl95}, bulk nuclear magnetic resonance \cite{gershenfeld:sci97}, low-capacitance Josephson junctions \cite{shnirman:prl97}, donor atoms \cite{kane:nat98, vrijen:pra00}, linear optics \cite{knill:nat01}, molecular magnets \cite{leuenberger:nat01},  spin clusters \cite{meier:prl03}, or color centers in diamond \cite{jelezko:pssa06, hanson:nat08, maletinsky:natnano12}. A long list of interesting results has followed, some of which will be reviewed in this article. 

\begin{figure}[htb]
\begin{center}
\includegraphics[width=1.00\linewidth]{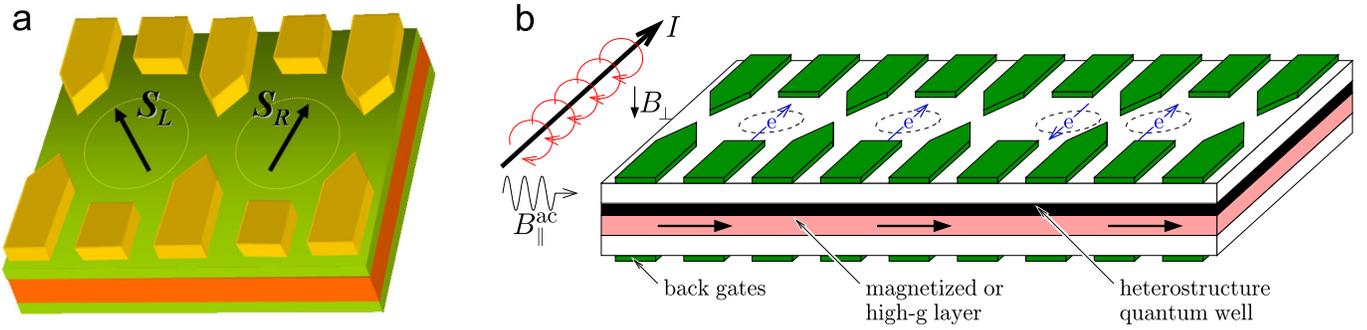}
\caption{Basic scheme for the physical implementation of a quantum computer as proposed in Ref.\ \cite{loss:pra98}. (a) The qubits are encoded in the spin states of single-electron quantum dots (QDs), where the barrier between adjacent QDs is controllable via electric gates. When the barrier is reduced, the electron wave functions overlap and the spins interact via the Heisenberg exchange coupling $J(t) \bm{S}_L \cdot \bm{S}_R$, where $J(t)$ is a function of gate voltage and corresponds to the energy splitting of the spin singlet and triplet states. This allows for electrically controlled two-qubit gates with fast operation times \cite{loss:pra98}. (b) A QD array based on (a), where the qubits in the two right dots are coupled while the others are decoupled. Rotations of individual spins may be achieved by pulling the desired electron down into a region of high magnetization or high $g$ factor via back gates, such that the Zeeman splitting and hence the resonance condition changes for this electron in the presence of a static magnetic field $B_\perp$. A resonantly applied oscillating magnetic field pulse $B_\parallel^{\rm ac}$ hence rotates the addressed qubit (electron spin resonance, ESR), while all others remain unaffected due to off-resonance. Exploiting spin-orbit interaction, the rotations may also be driven fully electrically via electric-dipole-induced spin resonance (EDSR). For details on ESR and EDSR, see Sec.\ \ref{sec:SOIAndNuclearSpins}. Alternatively, fast single-qubit gates may be implemented via exchange-controlled spin rotations \cite{loss:pra98, coish:prb07}. Combination of single- and two-qubit gates results in a universal set of quantum gates, so that the proposed schemes allow for fast and purely electrically controlled quantum computation with electron spins in QDs \cite{loss:pra98}.}
\label{schemeLDiVQC}
\end{center}
\end{figure}

In 1997, it was proposed to encode the quantum information in the spin states of single-electron quantum dots \cite{loss:pra98}. The tunnel barrier between neighboring dots, which can be varied via gates, Fig.\ \ref{schemeLDiVQC}, induces time-dependent electron-electron interactions and affects the spin states via the Heisenberg exchange coupling. The proposal demonstrates theoretically that such a setup allows for universal and scalable quantum computing, controllable by purely electrical means. Here, we will particularly focus on the experimental and theoretical achievements following Ref.\ \cite{loss:pra98}, since substantial progress in this field has been made within the past few years. An overview on recent results in other setups can, e.g., be found in Ref.\ \cite{ladd:nat10}.

The review is organized as follows. In Sec.\ \ref{sec:SystemsAndDefinitions} we introduce the three quantum dot (QD) systems discussed in this report: self-assembled QDs, lateral QDs, and nanowire-based QDs. We also comment on basic requirements for quantum computation and define the spin lifetimes $T_1$, $T_2$, and $T_2^{*}$. Spin-orbit interaction and nuclear spins are covered in Sec.\ \ref{sec:SOIAndNuclearSpins}. These present an undesired (noise) source of relaxation and decoherence for the spin qubits on the one hand, but on the other hand allow for all-electrical spin manipulation via electric-dipole-induced spin resonance, or for strong (effective) magnetic field gradients. In the main part, Sec.\ \ref{sec:RecentProgress}, recent progress in these QD systems is summarized, compared, and discussed. Newly proposed architectures for long-distance qubit-qubit coupling are reviewed in Sec.\ \ref{sec:NewProposals}, followed by our summary and final remarks, Sec.\ \ref{sec:Outlook}.

\section{Promising Quantum Dot Structures, Definition of Lifetimes, and Essential Requirements}
\label{sec:SystemsAndDefinitions}

Quantum dots (QDs) confine electrons or holes (missing valence band electrons) in all three dimensions, on length scales which are comparable to the wavelengths of the particles, i.e., typically $\sim \mbox{10-100 nm}$ in each spatial direction. There are many possibilities to realize such confinement, which is evident from the variety of systems under study. In this report, we will mainly focus on three of them. The first category is self-assembled QDs. These form naturally during growth, where InGaAs dots within a GaAs matrix are commonly used examples. When InGaAs is grown on GaAs (Stranski-Krastanov mode), islands form spontaneously after a small critical thickness of only a few monolayers due to the mismatch in the lattice constants. These may then be covered with further layers of GaAs. Such QDs are typically lens-shaped, with heights of $\sim \mbox{5 nm}$ (growth direction) and diameters $\sim \mbox{20 nm}$, and confinement results from the difference in the conduction and valence band edges of the involved materials. Alternatively, interface fluctuation QDs arise from monolayer fluctuations in thin quantum wells, typically resulting in GaAs dots within AlGaAs \cite{gywat:book}. The second category, lateral QDs, is based on two-dimensional electron and hole gases (2DEGs, 2DHGs), which exist in heterostructures from materials with suitable band properties and additional dopants. For instance, AlGaAs/GaAs heterostructures are routinely used to form 2DEGs and 2DHGs within GaAs, strongly confined along the growth direction. Lithographically defined gate electrodes on the sample surface allow to create confinement also in the transverse directions, leading to quasi-2D QDs of $\sim \mbox{100 nm}$ in diameter. Finally, semiconductor nanowires naturally provide confinement in two dimensions, due to their small diameters of $\sim \mbox{5-100 nm}$, and repulsive forces along the wire may again be added via nearby gates or via additional layers of barrier material. We note that several other QD implementations exist, which, however, for space reasons will not be discussed here. Prominent examples are QDs in carbon-based systems, like graphene \cite{trauzettel:nph07} or carbon nanotubes \cite{bulaev:prb08, klinovaja:prb11}, which are highly attractive for implementing spin qubits.

Any setup considered for quantum computation should fulfill a list of essential criteria, such as scalability and the ability to initialize the system in a fiducial state \cite{divincenzo:fph00}. For quantum error correction schemes to be applicable, it is important that the lifetimes are much longer than the gate operation times. A decade ago, this had been considered a very serious challenge, by now this problem seems to have been overcome in QDs as shown in Sec.\ \ref{sec:RecentProgress}. Three time scales are of interest in this context, which we illustrate in terms of the electron spin qubit states $\ket{\uparrow}$ and $\ket{\downarrow}$, assuming that these are eigenstates of the Pauli operator $\sigma_z$ with energy difference $\Delta$. First, the relaxation time $T_1$ describes transitions $\ket{\uparrow} \to \ket{\downarrow}$ due to interactions with the environment, such as the lattice, which leads to relaxation from the excited $\ket{\uparrow}$ to the ground state $\ket{\downarrow}$. A typical measure for $T_1$ is $\langle \sigma_z \rangle(t)$ with initial state $\ket{\uparrow}$. Second, the decoherence time $T_2$ quantifies the decay of quantum mechanical superpositions, and accounts for transitions of type $\ket{\uparrow} + \ket{\downarrow} \to \{\ket{\downarrow}, \ket{\uparrow}\}$ induced by the environment (see inset to Fig.\ \ref{figureDecoherenceScience}). When the state is initially $\left(\ket{\uparrow} + \ket{\downarrow} \right)/ \sqrt{2}$, an eigenstate of $\sigma_x$, a typical measure for $T_2$ is the envelope function of $\langle \sigma_x \rangle(t)$. We note that $\langle \sigma_x \rangle(t)$ oscillates between $\pm 1$ at angular frequency $\Delta/ \hbar$ for a perfectly isolated system, but decays to 0 as the state turns into either $\ket{\uparrow}$ or $\ket{\downarrow}$. The envelope function itself may be referred to as $\left|\langle \sigma_+ \rangle\right|(t) = \left|\langle \sigma_x \rangle(t) + i \langle \sigma_y \rangle(t) \right|$. Finally, in practice it is generally required to average over an ensemble, rather than to measure a single system only. The averaged $\left|\langle \sigma_+ \rangle\right|(t)$ often decays faster than in each individual case, because the oscillation frequencies may be slightly different from system to system (e.g., small deviations in $\Delta$), which leads to destructive interference and additional damping. The so-called dephasing time obtained from an ensemble measurement is therefore labeled $T_2^*$. The three time scales $T_1$, $T_2$, and $T_2^*$ are not completely unrelated. One finds that $T_2 \leq 2 T_1$, and usually $T_2^* < T_2$ and $T_2 \ll T_1$ in practice. In QDs, decoherence typically results from the nuclear spins and also the spin-orbit interaction. Commenting on the terminology, the relaxation ($T_1$), decoherence ($T_2$), and dephasing ($T_2^*$) times are only well defined when $\langle \sigma_z \rangle$ or $\left|\langle \sigma_+ \rangle\right|$, respectively, decay exponentially, which is the assumed behavior in most quantum error correction schemes. Strictly speaking, one should therefore avoid these terms when the longitudinal or transverse decay is of a non-exponential form. We note, however, that the introduced nomenclature is often being used to characterize any decaying behavior for convenience.  

A key criterion for building quantum computers, the one which actually justifies their name, is the presence of a universal set of quantum gates. This may fortunately be realized with one- and two-body interactions only, since any operation can be carried out as a sequence of one- and two-qubit gates. In fact, the implementation of single-qubit rotations for each element, along with only one type of entangling two-qubit gates, e.g., $\sqrt{\mbox{SWAP}}$ or cNOT, between neighboring qubits would be sufficient for universal quantum computation \cite{divincenzo:fph00, barenco:pra95}. We note in passing that entanglement of spin qubits can be created in many different ways and over long distances, for instance, by extracting and separating Cooper pairs from an s-wave superconductor as proposed in Ref.\ \cite{recher:prb01} and experimentally investigated recently in Refs.\ \cite{hofstetter:nat09, herrmann:prl10}. Nuclear spins and spin-orbit interaction, which present an undesired source of decoherence on the one hand, may prove useful for implementing qubit gates on the other hand, and both mechanisms will therefore be analyzed in more detail in the next section.

\section{Spin-Orbit Interaction and Nuclear Spins in Quantum Dots}
\label{sec:SOIAndNuclearSpins}

\subsection{Spin-Orbit Interaction}

Spin-orbit interaction (SOI) couples the orbital motion of a charge to its spin and arises in the presence of inversion asymmetry, where two types can be distinguished.  Rashba SOI results from structural inversion asymmetry and, for electrons, is typically of the form $H_R \propto \bm{E}_{\rm eff} \cdot \left(\bm{\sigma} \times \bm{p} \right)$, where the components $\sigma_i$ are the standard Pauli matrices for spin 1/2, $\bm{p}$ is the momentum operator, and $\bm{E}_{\rm eff}$ is an effective electric field determined by the system structure \cite{winkler:book}. Dresselhaus SOI is present in materials that lack bulk inversion symmetry, such as InAs or GaAs, and is of the form $H_D \propto p_{x^\prime}(p_{y^\prime}^2 - p_{z^\prime}^2)\sigma_{x^\prime} + p_{y^\prime}(p_{z^\prime}^2 - p_{x^\prime}^2)\sigma_{y^\prime} + p_{z^\prime}(p_{x^\prime}^2 - p_{y^\prime}^2)\sigma_{z^\prime}$, where $x^\prime$, $y^\prime$, and $z^\prime$ correspond to the crystallographic directions $[100]$, $[010]$, and $[001]$, respectively \cite{winkler:book, hanson:rmp07}. 

For quasi-2D systems, these Hamiltonians can be reduced further. For strong confinement along the $z$ direction, the Rashba term simplifies to $H_R = \alpha (p_x \sigma_y - p_y \sigma_x)$ with Rashba parameter $\alpha$. The resulting form of the Dresselhaus term strongly depends on the growth direction. 
For the $z$ axis chosen along the confinement direction, one can substitute $p_z \to \langle p_z \rangle = 0$, $p_z^2 \to \langle p_z^2 \rangle$, and all other terms can be neglected because of their smallness compared to terms $\propto\langle p_z^2 \rangle$ \cite{hanson:rmp07}. For example, for $z \parallel [100]$ one obtains $H_D = \beta (p_y \sigma_y - p_x \sigma_x)$, while for $z \parallel [110]$ the spin projection along the confinement direction is conserved, $H_D \propto p_x \sigma_z$. Both these Hamiltonians vary under rotations of the coordinate system about the $z$ axis, so that their exact form is determined by the relative orientation of coordinate and crystal axes. This is different for $z \parallel [111]$, where the effective Dresselhaus term is $H_D \propto p_x \sigma_y - p_y \sigma_x$, which moreover corresponds exactly to the form of the Rashba term. Therefore, Rashba and Dresselhaus SOI can cancel in lowest order for growth along the $[111]$ direction \cite{balocchi:prl11}. 

We note that the presence of SOI results in small, but finite, anisotropic corrections to the Heisenberg exchange interaction of electron spins, thus affecting the fidelity of quantum gates based on isotropic exchange. Fortunately, strategies have been developed with which the SOI-induced gate errors can be strongly suppressed \cite{bonesteel:prl01, burkard:prl02, stepanenko:prb03}. In general, gate errors can be reduced from first to second (or higher) order in SOI when the coupling strength $J(t)$ is varied symmetrically in time, followed by additional qubit rotations \cite{bonesteel:prl01, stepanenko:prb03}. In particular, the anisotropic corrections can be cancelled completely in the cNOT gate construction of Ref.\ \cite{loss:pra98} when the system is pulsed such that the anisotropic terms are linear in $J(t)$ \cite{burkard:prl02}. Additional errors caused by dipole-dipole interactions were found to be negligible for cNOT (in typical situations) \cite{burkard:prl02}.

\subsubsection{Electric-Dipole-Induced Spin Resonance}
\label{secsubsub:ESRandEDSR}

A rather useful technique for electrically controlled qubit rotations is the electric-dipole-induced spin resonance (EDSR). It is closely related to the well-known electron spin resonance (ESR), which we therefore review first. For this, let us consider an electron in a QD in the presence of magnetic fields. The Hamiltonian $H = H_0 + H_Z$ consists of a spin-independent part $H_0 = p^2/(2 m^*) + V(\bm{r})$, where the first (second) term corresponds to the kinetic (potential) energy, and the Zeeman part $H_Z = g \mu_B \bm{B} \cdot \bm{\sigma}/2$ which couples the magnetic field $\bm{B}$ to the spin. In the following, we assume that a constant magnetic field $B_z$ is applied along the $z$ axis, while a small oscillating field $B_x(t) = B_\perp \cos(\omega t)$, $B_\perp < B_z$, is applied along the $x$ axis. For any fixed orbital state $\ket{n}$, with $H_0\ket{n} = E_n \ket{n}$, the time evolution of the spin is described by the (von Neumann) master equation for the density matrix $\rho$,
\begin{gather}
\frac{d}{dt}\rho = -\frac{i}{\hbar}[E_n \bm{1}_2 + H_Z,\rho] = -\frac{i}{\hbar}[H_Z,\rho], \\
H_Z = \frac{\hbar \omega_z}{2} \left( \begin{array}{cc} 1 & 0 \\ 0 & -1 \end{array} \right) + \frac{\hbar \omega_\perp}{4} \left( \begin{array}{cc} 0 & e^{- i \omega t} \\ e^{i \omega t} & 0 \end{array} \right) + \frac{\hbar \omega_\perp}{4} \left( \begin{array}{cc} 0 & e^{i \omega t} \\ e^{-i \omega t} & 0 \end{array} \right),
\label{ZeemanTermExplicitForm}
\end{gather}   
where we defined $\hbar \omega_z \equiv g \mu_B B_z$, $\hbar \omega_\perp \equiv g \mu_B B_\perp$, and the states of the matrices correspond to $\{\ket{n,\uparrow}, \ket{n,\downarrow}\} \equiv \{\ket{\uparrow}, \ket{\downarrow}\}$. When $\omega \approx \omega_z$, the final term of the chosen representation of $\sigma_x \cos{\omega t}$, Eq.\ (\ref{ZeemanTermExplicitForm}), can be omitted because it only superimposes a fast and negligibly small oscillation to the dynamics. Within this rotating wave approximation, the resulting set of differential equations is exactly solvable. When the spin is originally in the $\ket{\uparrow}$ state and the oscillating field is applied for $t \geq 0$, the probability $p_\downarrow$ of a spin flip oscillates according to
\begin{equation}
p_\downarrow = \frac{\omega_\perp^2}{\omega_\perp^2 + 4 \delta^2} \sin^2\left(\frac{t}{4} \sqrt{\omega_\perp^2 + 4 \delta^2} \right),
\label{ESRresult}
\end{equation}
where $\delta \equiv \omega - \omega_z$ is the detuning from the resonance condition $\omega = \omega_z$, and $4 \pi /\sqrt{\omega_\perp^2 + 4 \delta^2}$ is the cycle duration. We note that the resulting spin-flip probability is completely equivalent to Eq.\ (\ref{ESRresult}) when the spin is initially down.

Remarkably, in the presence of SOI one finds that an oscillating electric field $\bm{E}(t) = \bm{E}_0 \cos(\omega t)$ leads to an effective magnetic field $\bm{b}_0 \cos(\omega t)$ with, in general, non-zero components perpendicular to the static magnetic field. Hence, spin rotations can efficiently be driven by purely electrical means (EDSR). This may be achieved by applying ac voltages to nearby gates, at frequencies which are in resonance with the Zeeman splitting, as recently exploited in experiments on nanowire-based InAs and InSb QDs \cite{nadjperge:nat10, schroer:prl11, nadjperge:arX12}. Explicit expressions for $\bm{b}_0$ are lengthy, and in the following we therefore comment on important properties found in an analysis for lateral QDs with growth axis $z \parallel [100]$ and harmonic confinement in the $x$-$y$ plane \cite{golovach:prb06}. First, in contrast to ESR, the EDSR arises from coupling to other orbital states and therefore depends on the level spacing. This can easily be seen, since $\bra{n}p_x\ket{n} = \bra{n}p_y\ket{n} = \bra{n}x\ket{n} = \bra{n}y\ket{n} = 0$, so that neither the dipolar term $e \bm{E}(t)\cdot \bm{r}$ nor $H_R$ or $H_D$ couple the spin states in lowest order. A unitary Schrieffer-Wolff transformation shows that the leading term for EDSR is a combination of Zeeman coupling and SOI. More precisely, the effective magnetic field in the ground state is $\propto \bm{B}_0 \times \bm{\Omega}(t)$, where $\bm{B}_0$ is the static magnetic field and $\bm{\Omega}(t) = \bm{\Omega}_0 \cos(\omega t)$ depends linearly on the electric field components in the $x$-$y$ plane and the parameters $\alpha$ and $\beta$ \cite{golovach:prb06}. We note that the resulting magnetic field is fully transverse and therefore most efficient. It can be quenched if $\bm{B}_0 \parallel \bm{\Omega}_0$. For typical GaAs QDs, EDSR allows spin manipulation on a time scale of 10 ns with the current experimental setups \cite{golovach:prb06}. An analysis for heavy-hole QDs can be found in Ref.\ \cite{bulaev:prl07}.

\subsubsection{Relaxation and Decoherence}

SOI, in combination with the phonon field, also leads to spin relaxation, characterized by the relaxation time $T_1$. For electrons in $[100]$-grown 2D QDs it has been calculated that this phonon-mediated mechanism results in $T_1 \propto \left(\hbar \omega_0\right)^4 / \left(g \mu_B B\right)^5 $ at low temperatures, where $g \mu_B B$ is the Zeeman splitting induced by a magnetic field $B$, and $\hbar \omega_0$ is the orbital level spacing \cite{khaetskii:prb01}. For moderate magnetic fields, this dependence agrees very well with experimental results \cite{hanson:rmp07, kroutvar:nat04, dreiser:prb08, amasha:prl08}. As the magnetic fields become very large, the wavelengths of the phonons with energy $g \mu_B B$ eventually become much smaller than the size of the QD, i.e., the phonon-induced effects average out when integrating over the electron wave function and $T_1$ increases rather than decreases \cite{golovach:prl04}. Maximal relaxation rates are usually observed when the phonon wavelength matches the dot size \cite{hanson:rmp07}. In the other limit, for very small magnetic fields, the derived expression for $T_1$ diverges. This is because the above theory focuses on single-phonon processes, so that only phonons in resonance with the Zeeman transition can contribute. Kramers' theorem forbids SOI-induced spin relaxation in the absence of a magnetic field \cite{hanson:rmp07, khaetskii:prb01}, which is also the reason why EDSR requires the presence of a finite magnetic field. When two-phonon processes are included, $T_1$ converges to a finite value \cite{abrahams:pr57, khaetskii:prb01, trif:prl09}.

Holes are an attractive alternative to electrons because of the suppressed contact hyperfine interaction with nuclear spins (see subsection below). Phonon-mediated spin relaxation has also been analyzed in detail for flat $[100]$-grown heavy-hole QDs with magnetic fields along the confinement axis \cite{bulaev:prl05}. For low magnetic fields, one finds $T_1 \propto B^{-5}$ due to Dresselhaus SOI, which is the same dependence as in the electron case, while the contribution due to Rashba SOI is $T_1 \propto B^{-9}$. The analysis shows that the spin relaxation time for heavy-holes can be comparable to or even longer than that for electrons when the QD is strongly 2D, illustrating that holes are very sensitive to confinement \cite{bulaev:prl05}. For instance, $T_1 > 0.2\mbox{ ms}$ has been measured for heavy-holes in self-assembled InGaAs QDs \cite{heiss:prb07, gerardot:nat08, gerardot:book}. In the limit $B \to 0$, the relaxation times are determined by two-phonon processes. These have been included theoretically \cite{trif:prl09}, suggesting times $T_1$ on the order of milliseconds, in good agreement with values observed in experiments \cite{gerardot:nat08, gerardot:book}. 

Notably, it has been shown that the upper limit $T_2 = 2 T_1$ is fulfilled in both the electron \cite{golovach:prl04} and the hole \cite{bulaev:prl05} cases discussed above, in contrast to the naively expected relation $T_2 \ll T_1$. Furthermore, theory predicts that electron spin relaxation is drastically suppressed for a certain magnetic field direction when $|\alpha| = |\beta|$ in $[100]$-grown QDs \cite{golovach:prl04}. [In passing we note that in this special limit, a new symmetry in spin space emerges giving rise to interesting spintronics effects in quasi-2D systems \cite{schliemann:prl03, duckheim:prb09}.] Tuning the Rashba coefficient via electric fields, this effect should be observable in an experiment with a vector magnet. The analysis of SOI-mediated relaxation has been extended to QDs with two electrons, forming spin singlet and triplet states, where magnetic field orientations with strongly suppressed spin relaxation were found to exist for arbitrary Rashba and Dresselhaus coefficients \cite{golovach:prb08}. The (relative) strengths of $\alpha$ and $\beta$ may be found via the singlet-triplet anticrossings for magnetic fields applied in growth direction, or by measuring the ``magic angles'' at which the singlet-triplet anticrossings and thus the corresponding singlet-triplet relaxations vanish in leading order of the SOI \cite{golovach:prb08, fasth:prl07}. Recently, a formula has been derived for lateral double quantum dots (DQDs), which quantifies the level splitting at the singlet-triplet anticrossing in terms of various parameters \cite{stepanenko:prb12}. This formula should allow to extract both the spin-orbit parameters and also the hyperfine coupling from transport or charge sensing experiments in such DQDs \cite{stepanenko:prb12}. Effects of hyperfine interaction will be discussed in following.

\subsection{Nuclear Spins}

A QD typically consists of $10^4$-$10^6$ atoms, so that an electron or hole confined to the QD overlaps with a large number of nuclear spins. The nuclear spin bath itself reveals large lifetimes, indicated by the long dipole-dipole correlation time among nuclear spins, $\sim$ 0.1 ms in GaAs \cite{koppens:prl07, coish:prb04}, but presents the main source of electron and hole spin decoherence. This is due to the hyperfine interaction among electron and nuclear spins, for which three different mechanisms can be derived from the Dirac equation \cite{fischer:ssc09}. The first one is the (isotropic) contact hyperfine interaction, which is the most relevant mechanism for conduction band electrons. For holes, where the Bloch functions are $p$- as opposed to $s$-type, the anisotropic hyperfine interaction and coupling to the orbital angular momentum become dominant instead. Below we summarize the effects and opportunities of a nuclear spin bath in more detail.  

\subsubsection{Electron Spin Decoherence}
\label{secsubsub:ElectronSpinDecoherence}

When we assume that the external magnetic field, if present, is oriented along the $z$ axis, the Hamiltonian of an electron spin $\hbar \bm{\sigma}/2$ coupled to a bath of nuclear spins $\hbar \bm{I}_k$ reads
\begin{equation}
\frac{1}{2}g \mu_B B_z \sigma_z + \frac{1}{2}\sum_k A_k \bm{I}_k \cdot \bm{\sigma} = \frac{1}{2}\Big(g \mu_B B_z + \sum_k A_k I_k^z \Big) \sigma_z + \frac{1}{4}\sum_k A_k \left(I_k^+ \sigma_- + I_k^- \sigma_+ \right),
\label{equationFlipFlop}
\end{equation}    
where $\bm{\sigma} = \left(\sigma_x, \sigma_y, \sigma_z\right)^{\rm T}$ is the vector of spin-1/2 Pauli matrices, $g$ is the effective electron $g$ factor, $B_z$ is the external magnetic field, $A_k$ (positive for In, Ga, As) are the contact hyperfine coupling constants, and $I_k^\pm = I_k^x \pm I_k^y$, $\sigma_\pm = \sigma_x \pm \sigma_y$, are the raising and lowering operators for nuclear and electron spin, respectively \cite{fischer:ssc09, coish:pssb09}. The effects of the nuclear spins on an electron spin in a QD can thus be described in terms of an effective magnetic field $\sum_k A_k \bm{I}_k / (g \mu_B)$, referred to as the Overhauser field. Its component $B_n^z$ along the $z$ axis changes the total Zeeman splitting by the Overhauser shift, while transverse components couple the spin states $\ket{\uparrow}$ and $\ket{\downarrow}$ through electron-nuclear-spin flip-flop processes \cite{coish:pssb09}. To avoid confusion, we mention that the term Overhauser field is typically used both for the 3D effective nuclear magnetic field and its component $B_n^z$. The largest possible value for $|B_n^z|$, obtained for a fully polarized bath, is $B_n^{\rm max} = I A/(|g| \mu_{\rm B})$, where $A$ is the averaged effective hyperfine coupling constant and $I$ is the (average) quantum number for the nuclear spin. For GaAs, $A \approx 90\mbox{ $\mu$eV}$ and $I = 3/2$, thus $B_n^{\rm max} \approx 5\mbox{ T}$ for the bulk $g$ factor -0.44, and we note that $B_n^{\rm max}$ is independent of the dot size \cite{fischer:ssc09, coish:pssb09}. Without further preparation, the $N$ nuclear spins inside a QD are in a superposition of states with different fields $B_n^z$, statistically distributed around a mean value $p B_n^{\rm max}$, where $-1 \leq p \leq 1$ is the nuclear spin polarization along the $z$ axis. Unless $|p|\to 1$, the width of this distribution is on the order of $B_n^{\rm max}/\sqrt{N}$, i.e., a few (tens of) mT for typical GaAs QDs of $\sim 10^4 \mbox{-} 10^6$ nuclear spins \cite{khaetskii:prl02, khaetskii:prb03, merkulov:prb02}. These internal fluctuations lead to dephasing and reduce the electron spin coherence time in GaAs dots to a few ns only \cite{khaetskii:prl02, khaetskii:prb03, merkulov:prb02, coish:prb04, fischer:ssc09, coish:pssb09}. The associated decay of the transverse spin is Gaussian and the decay time scales $\propto \sqrt{N}/ \big(A \sqrt{1- p^2} \big)$, when $I = 1/2$ and homogeneous coupling are assumed for simplicity \cite{coish:prb04}. One possibility to prolong the lifetimes, apart from increasing the dot size, is hence to polarize the nuclear spins, which will be discussed in more detail in paragraph \ref{secsubsub:DNP}. However, referring to the factor $\sqrt{1- p^2}$ obtained for $I= 1/2$, $|p| > 0.99$ is required to reduce the decoherence by a factor of ten.        

\begin{figure}[htb]
\begin{center}
\includegraphics[width=0.42\linewidth]{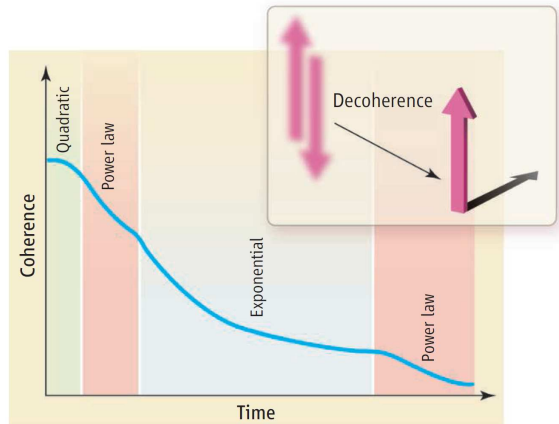}
\caption{Inset: Decoherence corresponds to the decay of quantum mechanical superpositions due to interaction with the environment. For the prominent example of an electron spin qubit with eigenstates $\ket{\uparrow}$, $\ket{\downarrow}$, decoherence refers to transitions of type ``$\ket{\uparrow}$ and $\ket{\downarrow}$'' $\to$ ``either $\ket{\uparrow}$ or $\ket{\downarrow}$'', and can be quantified by the time decay of the transverse spin $\left|\langle \sigma_+ \rangle\right|(t)$ as described in Sec.\ \ref{sec:SystemsAndDefinitions}. Main: Schematic decay of $\left|\langle \sigma_+ \rangle\right|(t)$ for a quantum dot electron with large Zeeman splitting, assuming that the nuclear spin bath has been prepared in a narrowed state and that echo pulses are absent. The sketch illustrates the variety of decay laws which the system proceeds through. We note that the initial quadratic decay occurs on an ultrashort time scale, while an additional quadratic shoulder appears at the transition from the power law to the exponential loss of coherence. Details can be found in Refs.\ \cite{coish:prb04, coish:pssb09, coish:prb10, coish:prb08, fischer:ssc09, fischer:sci09}. 
{\it The picture was taken from \cite{fischer:sci09} and is property of Science Magazine (www.sciencemag.org).} }
\label{figureDecoherenceScience}
\end{center}
\end{figure}

A second, attractive approach for lifetime prolongation is to narrow the intrinsic distribution for $|p|<1$ \cite{coish:prb04}. When the nuclear spin bath is initially in a less noisy, ``narrowed'' state, the electron spin decoherence induced by the finite width of possible $B_n^z$ is suppressed. In particular, this dephasing mechanism is overcome when the system is initially in an eigenstate to a field $B_n^z$. In this case, the decoherence time is no longer $\propto \sqrt{N}$, but $\propto N$, so that the coherence times can be increased by several orders of magnitude \cite{coish:prb04, fischer:ssc09, khaetskii:prl02, khaetskii:prb03, coish:pssb09, coish:prb08, coish:prb10}. 
The decay dynamics clearly differ from the Gaussian behavior which results from internal dephasing. In fact, an entire zoo of decoherence laws has been found, with a time decay that can proceed through several different stages, Fig.\ \ref{figureDecoherenceScience} \cite{fischer:sci09}. The reason for this feature is a rather long bath correlation time of order $\hbar N/A$ \cite{khaetskii:prl02, khaetskii:prb03, coish:pssb09, fischer:ssc09}. The dynamics of the isolated electron spin interacting with the nuclear bath are therefore history-dependent (non-Markovian), and a Markov approximation, for which the longitudinal and transverse spin components decay exponentially, is typically invalid. On time scales $< 0.1\mbox{ ms}$, where the dipolar coupling among nuclear spins can be ignored, this non-Markovian decay has been analyzed in great detail, and we list a few key results in the following. Assuming that the externally induced Zeeman splitting is large, such that $\left|g \mu_{\rm B} B_z\right| > A$ for $I$ of order unity, i.e., $\left|B_z\right| \gtrsim 3.5 \mbox{ T}$ for GaAs, direct electron spin flips are energetically forbidden, which gives rise to pure dephasing of the electron spin \cite{coish:prb10}. Under these conditions (perturbative approach possible), the various stages passed through by the transverse electron spin dynamics have been calculated with one unified and systematic method based on expansion of a generalized master equation \cite{coish:prb04, coish:prb08, coish:pssb09, fischer:ssc09, coish:prb10}. These stages include an ultrashort quadratic decay and an initial (partial) power law decay, followed by a quadratic shoulder, a dominant exponential decay, and a long-time power law decay. However, the exact behavior depends on various parameters, such as the QD dimensionality \cite{coish:pssb09, coish:prb10}. The Markovian regime, which gives rise to the exponential decay, is reached for sufficiently large $B_z$, and analytic expressions for the decoherence time $T_2 \propto N B_z^2$ have been found \cite{coish:prb08, fischer:ssc09, coish:prb10, cywinski:prb09}. This analysis was also of interest from a technical point of view, since it verified that calculations based on high-order expansions of a leading-order effective Hamiltonian can have limited validity. In particular, a notable modulation of the decay envelope found at long times cannot be obtained with an effective Hamiltonian (see Ref.\ \cite{coish:prb10} and references therein). Observing these additional oscillations experimentally would be a desirable confirmation of the theory. At low magnetic fields, an expansion of the generalized master equation is not possible, and the spin dynamics still are not understood in detail. A list of available approaches allowing for some insight into the low-field regime can be found in Refs.\ \cite{coish:pssb09, liu:advphys10}. For instance, the system is exactly solvable in the special case when the nuclear spin bath is initially fully aligned \cite{khaetskii:prl02, khaetskii:prb03}. Independently of $B_z$, it turns out that both $\langle \sigma_z \rangle(t)$ and $\left|\langle \sigma_+ \rangle\right|(t)$ first perform small oscillations due to coherent exchange with the nuclear spin bath. After the bath correlation time of order $\hbar N/A$, electron spin coherence is irreversibly lost and the spin components converge to constant values, slightly below the initial values only, where the system remains until dipole-dipole interactions among the nuclear spins become relevant \cite{khaetskii:prl02, khaetskii:prb03}. For $B_z = 0$, assuming a Gaussian envelope wave function, this asymptotic time decay evolves according to $\ln^{-3/2} (t^\prime)$ and $\ln^{-1} (t^\prime)$ for 3D and 2D QDs, respectively, where $t^\prime \propto tA/N$. At large $B_z$, this decay is $\propto (t^\prime)^{-3/2}$ and $\propto (t^\prime)^{-1}$ in 3D and 2D dots, in agreement with the systematic solutions of the generalized master equation \cite{coish:prb04, khaetskii:prl02, khaetskii:prb03}.

\subsubsection{Hole Spin Decoherence}

The spin dynamics have also been investigated for heavy-holes in quasi-2D QDs \cite{fischer:ssc09, fischer:prb08, fischer:prl10, maier:arX12}. Assuming the strong confinement along the $z$ axis, we recall that the states can be classified according to their angular momenta (effective spins) $\ket{J, m_J}$, where $J$ quantifies the size and $m_J \in \{-J, -J+1,... ,J \}$ is the $z$-projection in units of $\hbar$. This results from the strong SOI in the valence band, coupling the electron spin to the $p$-type Bloch functions. Heavy-holes (HHs) have $\ket{3/2, \pm 3/2}$, while light-hole (LH) states have $\ket{3/2, \pm 1/2}$, and the two bands are energetically well separated in 2D-like QDs. Even though the contact hyperfine term is absent, the remaining mechanisms, i.e., anisotropic hyperfine interaction and coupling to the orbital angular momentum, turn out to be rather strong, in typical III-V compounds only one order of magnitude weaker than the contact hyperfine interaction for electrons \cite{fischer:prb08, eble:prl09, testelin:prb09, fallahi:prl10, chekhovich:prl11}. Remarkably, the coupling of the HH to the nuclear spin bath takes on a simple Ising form in leading order, $\sum_k A_k^h I_k^z s_z$, where $A_k^h$ is the coupling to the $k$th nucleus, $I_k^z$ denotes the $z$ component of the $k$th nuclear spin in units of $\hbar$, and $s_z$ is the HH pseudospin operator with eigenvalues $s_z = \pm 1/2$ for $m_J = \pm 3/2$ \cite{fischer:prb08, maier:arX12}. This clearly differs from the Heisenberg-type contact hyperfine interaction $\sum_k A_k \bm{I}_k\cdot \bm{\sigma}/2$, Eq.\ (\ref{equationFlipFlop}), because transverse components are basically absent in the case of holes.  

As for electrons, one can distinguish between different initial configurations. In the case of an unprepared, inhomogeneously broadened nuclear spin bath, the transverse hole spin decay depends on the orientation of the external magnetic field $\bm{B}$. For zero field or $\bm{B} \parallel z$, dephasing results in a Gaussian time decay, as for electrons, with time scales of typically a few tens of ns \cite{fischer:prb08}. However, due to the Ising-type rather than Heisenberg-type HH-nuclear-spin interaction, the situation changes drastically for an in-plane magnetic field $\bm{B} \parallel x$. Since the hyperfine fluctuations are now purely perpendicular to the applied field, one finds that the transverse hole spin decays $\propto \sqrt{E_{Z,x}/(\langle E_{n,z}^2\rangle t)}$ at long times in the limit  $E_{Z,x}^2 \gg \langle E_{n,z}^2\rangle$, where $E_{Z,x} = \left|g_x \mu_B B\right|$ is the externally induced Zeeman splitting, $g_x$ is the in-plane HH $g$ factor, and $\langle E_{n,z}^2\rangle$ is the variance of the nuclear field $\sum_k A_k^h I_k^z$ \cite{fischer:prb08, fischer:ssc09}. For typical GaAs QDs and magnetic fields of a few Tesla, the associated decay times are long, around tens of $\mu$s \cite{fischer:prb08}. Only a few months after these calculations were published, an experiment on self-assembled InGaAs QDs with $\bm{B} \parallel x$ confirmed that HH spins in 2D-like QDs are highly coherent, with $T_2^{*}> 0.1\mbox{ $\mu$s}$ ($T_2^{*} > 1\mbox{ $\mu$s}$ with $\sim 40\%$ probability) reported for the setup under study \cite{brunner:sci09, gerardot:book}. 

For applications which require large Zeeman splittings, an in-plane magnetic field may be inconvenient because $g_x$ is usually much smaller than the HH $g$ factor $g_z$ along the axis of strong confinement. Long coherence times for $\bm{B} \parallel z$ can be achieved as well, namely by preparing the nuclear spin bath in a narrowed state \cite{fischer:prl10, maier:arX12}. When the nuclear spins are initially in an eigenstate of $\sum_k A_k^h I_k^z$, with $\bm{B} \parallel z$, decoherence can only result from additional transverse terms in the HH-nuclear-spin coupling which then allow for flip-flop processes. These additional terms mainly arise from coupling to neighboring bands, i.e., the conduction band, LH band, and split-off band, and are about one to two orders of magnitude weaker than the dominant Ising term \cite{fischer:prl10, maier:arX12}. It turns out that the time decay of the transverse spin due to band hybridization is purely exponential, and that the decoherence time $T_2$ can be tuned over several orders of magnitude via the applied magnetic field. In fact, decoherence due to nuclear spins can be so strongly suppressed that other mechanisms, like the dipole-dipole interaction or the phonon bath, should take over as the dominant sources of transverse spin decay \cite{fischer:prl10, maier:arX12}. Calculations on self-assembled QDs showed that the Ising-like form of the hyperfine coupling is preserved for realistic strain distributions \cite{maier:arX12}. The strain considerably affects the hyperfine-induced hole spin decoherence, largely through coupling to the conduction band, allowing to tune $T_2$ by an order of magnitude for fixed Zeeman splittings \cite{maier:arX12}.

\subsubsection{Distribution Narrowing and Dynamic Nuclear Polarization}
\label{secsubsub:DNP}

As summarized above, the electron and hole spin decoherence induced by nuclear spins can be strongly suppressed when the nuclear spin bath is initialized in a narrowed state. Moreover, the effective nuclear magnetic field, up to $\sim\pm\mbox{5 T}$ for electrons in GaAs QDs, allows to realize large magnetic field gradients among neighboring QDs \cite{foletti:nph09} and to tune the resonance energies in optically active QDs over several tens of $\mu$eV \cite{kloeffel:prl11, latta:nph09, urbaszek:book, urbaszek:arX12}. Therefore, nuclear spins are more and more considered a source of opportunity rather than trouble, leading to enormous experimental efforts in this field. Dynamic nuclear polarization (DNP) schemes are usually based on the isotropic contact hyperfine interaction among electron spin and nuclear spins. As illustrated in paragraph \ref{secsubsub:ElectronSpinDecoherence}, Eq. (\ref{equationFlipFlop}), the transverse components of the Overhauser field allow to polarize the nuclei via electron-nuclear-spin flip-flop processes, building up a large nuclear field $\sum_k A_k I_k^z$ with maximum Overhauser shift $\left|g\right| \mu_B B_n^{\rm max} = A I$. An example for DNP via the hole spin can be found in Ref.\ \cite{xu:nat09}. 

DNP has been achieved experimentally by optical \cite{latta:nph09, kloeffel:prl11, urbaszek:book, urbaszek:arX12, xu:nat09, bracker:prl05, tartakovskii:prl07, maletinsky:prl07, makhonin:apl08, belhadj:prl09, chekhovich:prl10, klotz:prb10}, electrical \cite{petta:prl08, foletti:nph09, bluhm:prl10, barthel:arX11, laird:prl07, laird:sst09}, and also magnetical \cite{danon:prl09, vink:nph09} driving, where it is impossible to completely list the large variety of approaches. Polarizations $50\% < |p| < 70\%$ have been reported so far \cite{bracker:prl05, chekhovich:prl10, klotz:prb10}, however, achieving $|p|> 90\%$ remains a very challenging task. Interestingly, many DNP schemes feature an intrinsic feedback mechanism which drives the system towards fixed, stable nuclear field values, so that the width of the nuclear field distribution is narrowed at the same time \cite{latta:nph09, kloeffel:prl11, danon:prl09, vink:nph09, xu:nat09}. Similar effects have been demonstrated using pulsed optical excitation on an ensemble of QDs \cite{greilich:sci06, greilich:sci07}, and it has been shown that efficient feedback loops may also be included intentionally \cite{bluhm:prl10}. Further promising approaches for the preparation of narrowed states are based on indirect measurement \cite{stepanenko:prl06, klauser:prb06, giedke:pra06, barthel:prl09}. 

We mention that all decay properties described earlier in this section correspond to the free-induction decay, i.e., the case where the system evolves in the absence of externally applied control sequences. The dephasing due to inhomogeneous broadening can be undone to a large extent by applying spin-echo pulses, notably increasing the spin coherence times \cite{petta:sci05, koppens:prl08, greilich:nph09, press:nphot10, barthel:prl10, bluhm:nph11, medford:arX11}. In parallel, an alternative trend is to bypass the interaction with nuclear spins completely by switching to host materials such as Ge and Si, which can be grown nuclear-spin-free. Examples for both approaches will be discussed in the next section, where we review recent progress toward quantum computation with spins in QDs. Finally, we note that an alternative route to reduce the nuclear spin noise (besides narrowing and DNP) would be to polarize the nuclear spins by freezing them out, either by applying a sufficiently strong magnetic field on the order of $15\mbox{ T}$ at a few mK \cite{chesi:prl08} or by inducing an ordering transition of the nuclear spin system due to RKKY interactions. The latter phenomenon has attracted a lot of interest in recent years, and we refer the interested reader to the literature \cite{simon:prl07, braunecker:prb09, zak:prb12, governale:vpoint12}.

\section{Recent Progress In Quantum Dot Systems}
\label{sec:RecentProgress}

Since the first proposals in the 1990s, researchers worldwide have been working hard toward the ambitious goal of implementing a quantum computer. In quantum dots (QDs), which we focus on in this report, seminal progress has been made within the past few years. In the following, we summarize and discuss some of the key results, where we distinguish between self-assembled, lateral, and nanowire-based QDs. A table at the end of this section, Table \ref{TableComparison}, summarizes relevant information such as the measured lifetimes, and compares $T_2$ to reported operation times as commented below.  

Quantum systems are sensitive, and errors inevitably occur in any realistic device. Therefore, schemes for fault-tolerant quantum computing have been developed, where errors can automatically be corrected as long as they occur with low enough probability. The latter condition can usually be quantified in terms of a threshold rate \cite{aharonov:procstoc97, knill:sci98, preskill:prsla98}. For instance, standard error correction schemes require that at least $\sim 10^4$ gate operations can be carried out within the decoherence time of a qubit \cite{aharonov:procstoc97, knill:sci98, preskill:prsla98, awschalom:book, steane:pra03, aliferis:pra08}. A few years ago, a novel scheme has been presented, referred to as the surface code \cite{raussendorf:prl07, raussendorf:njp07, fowler:pra09, wang:pra11, divincenzo:physscr09}. The logical qubits are encoded within several physical qubits each, all of which are arranged in a 2D lattice with nearest neighbor interactions. Logical single- and two-qubit gates are then performed via a series of projective measurements. This code is probably the most powerful quantum computing scheme presently known, with a remarkably large error threshold around 1\% \cite{raussendorf:prl07, raussendorf:njp07, fowler:pra09, wang:pra11, divincenzo:physscr09}, so that $\sim 10^2$ operations per decoherence time may already be sufficient. When errors for readout become negligible, error correction in the surface code is even possible up to a threshold rate of currently 18.5\% \cite{wootton:arX12}. The viability of topological error correction has recently been demonstrated in a first proof-of-principle experiment \cite{yao:nat12}. For further information on fault-tolerant quantum computation with the surface code we refer to Refs.\ \cite{raussendorf:njp07, raussendorf:prl07, fowler:pra09, wang:pra11, divincenzo:physscr09, wootton:arX12, yao:nat12}. A general overview on fault-tolerance is provided in \cite{divincenzo:physscr09}.

\subsection{Self-Assembled Quantum Dots}
\label{secsub:ProgressSelfAssembled}

As opposed to gate-defined QDs, the confinement in self-assembled dots purely arises from the conduction and valence band offsets of the involved materials, leading to strong confinement on a very small scale in all three dimensions. Therefore, self-assembled QDs are typically operational at $4\mbox{ K}$, the boiling temperature of $^4$He. Moreover, they are optically active and feature strong interband transitions with ``almost hard'' selection rules \cite{gywat:book, gerardot:book, urbaszek:arX12}. Exploiting this property, self-assembled QDs are primarily studied using optical means, and heterostructures have been designed which allow precise control over the charge states \cite{drexler:prl94, warburton:nat00}.

For a qubit with basis $\ket{1}$ and $\ket{0}$, the general qubit state can be written as $\ket{\psi} = \cos(\theta/2)\ket{1} + e^{i \phi}\sin(\theta/2)\ket{0}$, neglecting global phases, where $0\leq \theta \leq \pi$ and $0\leq \phi < 2\pi$ correspond to the polar and azimuthal angle of a point on the Bloch sphere. Prerequisites for the implementation of a quantum computer are the abilities to initialize, to control, and to read out such a qubit state. All of this has been achieved by now. First, the spins of both electrons \cite{atatuere:sci06} and holes \cite{gerardot:nat08, godden:apl10} can be initialized with $\geq 99\%$ fidelity. Second, ultrashort optical pulses, combined with an externally induced Zeeman splitting, have allowed for complete quantum control, i.e., arbitrary rotations on the Bloch sphere, with operation times on the order of only a few ps. Again, this has successfully been demonstrated both on single electrons \cite{press:nphot10, berezovsky:sci08, press:nat08} and single holes \cite{degreve:nph11, greilich:arX11, godden:prl12}. Recently, initialization and coherent control have also been reported for two-particle qubits defined by the spin singlet and triplet states of electrons \cite{kim:nph11} and holes \cite{greilich:arX11} in vertically stacked QDs. The latter, also referred to as QD molecules, form naturally during growth when a second layer of QDs is grown on top of another. Even though the position of the dots in the first layer is arbitrary, the QDs in the second layer will form right on top of the first ones due to strain in the tunnel barrier grown inbetween. Finally, several methods have been developed to read out the spin states. These include time-averaged readout via Faraday rotations \cite{atatuere:nph07}, Kerr rotations \cite{berezovsky:sci06}, and resonance fluorescence \cite{vamivakas:nph09}. In addition, time-resolved Kerr rotation spectroscopy has been reported \cite{mikkelsen:nph07, berezovsky:sci08}, and it has been shown that QD molecules allow to measure the spin state of a single electron in real time via the resonance fluorescence \cite{vamivakas:nat10}. In the latter case, the presence of a second dot enables to use different optical transitions (laser energies) for the initialization and readout steps \cite{kim:prl08, vamivakas:nat10}.   

Dephasing and decoherence times have been measured. For single-electron spin qubits, dephasing results in $T_2^* \simeq\mbox{1-10 ns}$ \cite{greilich:sci06, press:nphot10, berezovsky:sci06, berezovsky:sci08, mikkelsen:nph07, xu:nph08} depending on the exact system, while the observed decoherence times are $T_2 \simeq\mbox{3 $\mu$s}$ \cite{greilich:sci06, press:nphot10}. However, it has been verified already that $T_2^*$ can be significantly prolonged by narrowing the distribution of the nuclear field \cite{xu:nat09, greilich:sci06, greilich:sci07}. For single holes, the reported $T_2^* = \mbox{2-21 ns}$ \cite{degreve:nph11, greilich:arX11, godden:prl12} and $T_2 =\mbox{1.1 $\mu$s}$ \cite{degreve:nph11} are similar to those for electrons, hence shorter than initially expected, which is attributed to electrical noise \cite{degreve:nph11, greilich:arX11}. Measurements based on coherent population trapping, i.e., in the frequency domain as opposed to the time domain, have revealed a much longer hole-spin dephasing time $> 0.1\mbox{ $\mu$s}$ \cite{brunner:sci09, gerardot:book}. Electrical noise is also the reason for the short $T_2^* = \mbox{0.4-0.7 ns}$ \cite{kim:nph11} and $T_2^* \leq \mbox{0.6 ns}$ \cite{greilich:arX11} observed for singlet-triplet qubits from electrons and holes, respectively. When the QD molecule is operated in a regime where the singlet-triplet splitting is less sensitive to fluctuating electric fields, $T_2^*$ can be increased by several orders of magnitude, and dephasing times up to \mbox{200 ns} have recently been measured for coupled electrons \cite{weiss:arX12}. Importantly, comparing the listed $T_2$ to the notably short operation times of a few (tens of) ps illustrates that the threshold factor of $\sim 10^4$ for standard quantum error correction schemes has already been exceeded both in electron and hole systems. For reported spin relaxation times $T_1$, see Table \ref{TableComparison}.

The optical properties make self-assembled QDs highly promising candidates for applications as single-photon sources, and designs for enhanced extraction efficiencies are being developed \cite{claudon:nphot10}. Moreover, they may present an interface between ``stationary'' and ``flying'' qubits. Recently, interference of single photons from two separate QDs has been demonstrated \cite{flagg:prl10, patel:nphot10}, which is a promising approach for generating entanglement and implementing two-qubit gates between distant spins. Since different QDs, unlike atoms, have different resonance energies, this requires the ability to tune the level structure of a dot for the photon energies to be matchable. Such fine tuning may be achieved via strain \cite{flagg:prl10, kuklewicz:arX12}, via the Stark effect \cite{patel:nphot10, kistner:optexpr08}, or by polarizing the nuclear spins up to a desired (Overhauser) nuclear magnetic field. Dynamic nuclear polarization in single self-assembled QDs has been studied in detail \cite{urbaszek:book, urbaszek:arX12, latta:nph09, kloeffel:prl11, xu:nat09, bracker:prl05, tartakovskii:prl07, maletinsky:prl07, makhonin:apl08, belhadj:prl09, chekhovich:prl10, klotz:prb10}. For instance, schemes exist both for high \cite{latta:nph09} and low \cite{kloeffel:prl11} external magnetic fields which allow for continuous, bidirectional tuning of the Overhauser field, fully controlled by the laser wavelength. Both schemes are also capable of narrowing the width of the nuclear field distribution \cite{latta:nph09, kloeffel:prl11}.

As an alternative to the widely studied III-V compounds, self-assembled QDs can also be grown from group IV materials. A prominent example are self-assembled Ge QDs on Si substrates. Due to the indirect band gap of Ge and Si, these dots are not as optically active as typical III-V QDs \cite{wang:procieee07}. Self-assembled Ge/Si QDs are subject to large experimental efforts, and detailed knowledge about growth and the electrical and optical properties has been gained \cite{wang:procieee07, katsaros:natnano10, katsaros:prl11}.

\subsection{Lateral Quantum Dots}

Substantial progress has been made on implementing gate-controlled qubits within the two-dimensional electron gas (2DEG) of a heterostructure. Experiments have predominantly been carried out on GaAs QDs within AlGaAs/GaAs heterostructures, which therefore should be considered to be the host material in the following, unless stated otherwise. Mainly, two different approaches for encoding the qubit have emerged. The first one follows the original proposal \cite{loss:pra98}, using the spin eigenstates $\ket{\uparrow}$ and $\ket{\downarrow}$ of single electrons. The second scheme uses the singlet $\ket{S} = \left(\ket{\uparrow \downarrow} - \ket{\downarrow \uparrow} \right)/\sqrt{2}$ and triplet $\ket{T_0} = \left(\ket{\uparrow \downarrow} + \ket{\downarrow \uparrow} \right)/\sqrt{2}$ states of two electron spins, forming an $S$-$T_0$ qubit \cite{levy:prl02, benjamin:pra01, taylor:nph05}. We summarize the results for both approaches in the following. Alternatively, $S$-$T_+$ qubits \cite{petta:sci10} and qubits from three-spin states \cite{divincenzo:nat00, laird:prb10, gaudreau:nph12, koppens:nph12} are currently under investigation.

\subsubsection{Two-Spin Qubits ($S$-$T_0$)}

In the $S$-$T_0$ approach, every qubit is formed by two electrons in two adjacent QDs, where the charge configuration and the overlap of the electron wave functions depend on the shape of the confining potential, which in turn is determined by the applied gate voltages. Within the frame of this review it is sufficient to distinguish the two charge configurations (1,1) and (0,2), which denotes that the electrons are found either in different dots or both in the ``right'' QD, respectively. Coherent rotations $\ket{S} \leftrightarrow \ket{T_0}$ are induced by a magnetic field gradient between the two QDs, typically achieved by dynamic polarization of the nuclear spins \cite{bluhm:prl10, foletti:nph09, barthel:arX11} or by a nearby positioned micromagnet \cite{laird:prl07, laird:sst09, pioroladriere:nph08, shin:prl10, brunner:prl11}. When the barrier is reduced such that the wave functions strongly overlap, the finite exchange energy gives rise to coherent rotations $\ket{\uparrow \downarrow} \leftrightarrow \ket{\downarrow \uparrow}$, i.e., $\left(\ket{T_0} + \ket{S} \right)/\sqrt{2} \leftrightarrow \left(\ket{T_0} - \ket{S} \right)/\sqrt{2}$, thus allowing for arbitrary rotations on the Bloch sphere \cite{foletti:nph09}. These single-qubit operations can be carried out within a few ns only \cite{bluhm:prl10, foletti:nph09, petta:sci05}, and two-qubit gates may be implemented by capacitive coupling, where the charge configuration in one double quantum dot (DQD) affects the exchange energy and hence the precession frequency in the other DQD \cite{taylor:nph05, vanweperen:prl11, shulman:arX12}. Charge-conditional phase flips of an $S$-$T_0$ qubit have recently been demonstrated in four-dot systems \cite{vanweperen:prl11, shulman:arX12}, and it was verified experimentally that the resulting cPHASE gate between two $S$-$T_0$ qubits is entangling \cite{shulman:arX12}.

For initialization, the potential energy in one of the dots is reduced such that both electrons tend to occupy the ground state of the same QD. This sets the qubit in the singlet state, because the symmetric orbital part of the two-electron wave function requires an antisymmetric contribution of the spin. After operation in the (1,1) regime, the same idea also allows for spin-to-charge conversion and therefore presents a popular basis for readout schemes. Having reduced the potential energy in one QD, the system changes to (0,2) for $\ket{S}$, but remains in (1,1) for $\ket{T_0}$ due to Pauli exclusion. The charge state can be detected via nearby quantum point contacts (QPCs) \cite{elzerman:prb03, elzerman:nat04}, which may furthermore be embedded in radio-frequency (rf) impedance matching circuits to allow for faster readout \cite{reilly:apl07, cassidy:apl07, barthel:prl09}. However, the quality of the outcome (readout visibility, see also Sec.\ \ref{secsub:OverviewProgress}) strongly depends on the triplet relaxation time at the measurement point, which may be clearly reduced in the presence of large magnetic field gradients \cite{barthel:arX11}, and implementing schemes for fast and reliable single-shot readout therefore remains an important task. Recently, single-shot measurements with measurement times down to 100 ns have been reported, using an rf sensor QD which is much more sensitive than standard QPCs \cite{barthel:prb10}. As an alternative approach, dispersive readout of spin singlet and triplet states has been demonstrated with an rf resonant circuit coupled to a DQD \cite{petersson:nanolett10}.  Also, readout via spin-dependent tunnel rates has been achieved for the singlet and triplet states in a single QD \cite{hanson:prl05, meunier:prb06, hanson:rmp07}.  

Even though the random distribution of the nuclear spins leads to rapid dephasing within $\sim 10\mbox{ ns}$ \cite{petta:sci05, bluhm:prl10, johnson:nat05}, two-electron spin states feature very long coherence times. A single echo pulse increases the dephasing time to a few $\mu$s already \cite{petta:sci05, barthel:prl10, bluhm:nph11, medford:arX11}, and more sophisticated pulse sequences have demonstrated coherence times on the order of a hundred $\mu$s \cite{barthel:prl10, medford:arX11, bluhm:nph11}, where $\mbox{276 $\mu$s}$ currently corresponds to the largest value reported so far \cite{bluhm:nph11}. Assuming that the two-qubit gates can be operated on a similar time scale (ns) as the single-qubit gates, the threshold value of $\sim 10^4$ operations per decoherence time is hence clearly exceeded. As we summarize in the following, this similarly holds for the case where the qubits are encoded in the spin states of single electrons.

\subsubsection{Single-Spin Qubits}   

The two eigenstates of a single electron in a QD, $\ket{\uparrow}$ and $\ket{\downarrow}$, are split by an effective Zeeman energy via coupling to external and internal (Overhauser) magnetic fields. This energy difference gives rise to coherent single-qubit rotations $\left(\ket{\uparrow} + \ket{\downarrow} \right)/\sqrt{2} \leftrightarrow \left(\ket{\uparrow} - \ket{\downarrow} \right)/\sqrt{2}$. Rotations about the second axis of the Bloch sphere, $\ket{\uparrow} \leftrightarrow \ket{\downarrow}$, can be driven by means of E(D)SR, which we introduced in Sec.\ \ref{sec:SOIAndNuclearSpins} of this review. Finally, two-qubit gates such as $\sqrt{\mbox{SWAP}}$ and SWAP can be implemented by controlling the overlap of the wave functions and hence the exchange energy for neighboring electrons. Thirteen years after publication of the original proposal \cite{loss:pra98}, the elementary unit of an all-electrical ``spin-qubit processor'' has now been implemented for the first time, demonstrating independently controllable single-spin rotations combined with inter-dot spin exchange in a DQD \cite{brunner:prl11}. An important part of the setup in Ref.\ \cite{brunner:prl11} is a micromagnet at the sample surface, whose stray field provides a time-independent magnetic field gradient at the QDs. This gradient is useful for two reasons. First, it allows for efficient and electrically driven ESR; an oscillating electric field slightly shifts the position of the electron in the QD, so that the electron ``feels'' an oscillating magnetic field of the same frequency without the need for SOI \cite{pioroladriere:nph08, brunner:prl11}. Second, it leads to shifted resonance frequencies in adjacent QDs via a difference in the Zeeman energies, so that neighboring qubits can be addressed individually \cite{laird:prl07, laird:sst09, pioroladriere:nph08, shin:prl10, brunner:prl11}.

Performance of the implemented spin-qubit processor was tested via a time-averaged readout scheme, using a nearby QPC to measure the average charge configuration when the system is tuned to the Pauli spin blockade regime (see also $S$-$T_0$ qubits above) \cite{brunner:prl11}. A fully operational unit for quantum computation requires precise initialization and single-shot readout of the individual qubit states, which therefore still needs to be included. Latest developments are promising \cite{shin:prl10, elzerman:nat04, barthel:prb10, nowack:sci11}, and independent single-shot readout of two electron spins in a DQD, with fidelities close to 90\%, has recently been reported \cite{nowack:sci11}. In the experiments of Ref.\ \cite{nowack:sci11}, a time around $1\mbox{ ms}$ was required for spin readout and initialization to $\ket{\uparrow}$, slightly shorter than the measured spin relaxation time of $\sim \mbox{4-5 ms}$. The latter strongly depends on the regime of operation, i.e., the applied gate voltages, and can take values $T_1 > \mbox{1 s}$ \cite{amasha:prl08}. The decoherence times $T_2$ can be assumed to be much longer than a $\mu$s. For a single echo pulse and a magnetic field of $\mbox{70 mT}$ only, a decay time near $\mbox{1 $\mu$s}$ has been measured, and clearly longer coherence times can be expected at higher magnetic fields and more sophisticated pulse sequences \cite{koppens:prl08}.

Comparing $T_2$ to the typical operation times, we find that the threshold rates of proposed (standard) error correction protocols have already come close within reach, as summarized in the following. Two-qubit gates are fast. For instance, SWAP operations $\ket{\uparrow \downarrow} \to \ket{\downarrow \uparrow}$ require a time of order $h/\left( 2 J\right)$, where $h$ is Planck's constant and $J$ is the exchange energy. They are routinely carried out within $\lesssim 10 \mbox{ ns}$ \cite{vanweperen:prl11, brunner:prl11}, and SWAP times $< 0.5 \mbox{ ns}$ have been demonstrated \cite{petta:sci05}. Furthermore, single-qubit rotations induced by the Zeeman splitting typically occur on a sub-ns time scale. Using the bulk $g$ factor $-0.44$ for GaAs and $B_z = 1\mbox{ T}$ as the magnetic field along the quantization axis, the $\pi$-rotation $\left(\ket{\uparrow} + \ket{\downarrow} \right)/\sqrt{2} \to \left(\ket{\uparrow} - \ket{\downarrow} \right)/\sqrt{2}$ requires a time of order $h/\left| 2 g \mu_B B_z\right| \simeq 0.1\mbox{ ns}$. The efficiency of the universal set of qubit gates is therefore limited by the operation time $\sim h/\left| g \mu_B B_\perp \right|$ for coherently driven spin flips $\ket{\uparrow} \to \ket{\downarrow}$, where $B_\perp$ is the amplitude of the oscillating magnetic field perpendicular to the quantization axis (see also paragraph \ref{secsubsub:ESRandEDSR}). Recently reported spin-flip times are on the order of several tens of ns, both for magnetically \cite{koppens:nat06, koppens:prl07, koppens:prl08} and electrically \cite{pioroladriere:nph08, nowack:sci07} driven rotations, corresponding to $B_\perp \approx \mbox{1-10 mT}$. The efficiency of electrically controlled rotations may, among others, be improved by increasing the magnetic field gradient \cite{pioroladriere:nph08}, or even, in a brute-force approach, by switching to host materials with stronger SOI \cite{nowack:sci07} and larger $g$ factors. Alternatively, coherent single-spin rotations about an arbitrary axis may be implemented using an auxiliary spin in a nearby QD with different Zeeman field, which enables a purely exchange-based control without the need for SOI or oscillating fields \cite{loss:pra98, coish:prb07}. Such exchange-controlled single-qubit gates have not yet been realized experimentally; theoretical results, however, are promising and point toward high fidelities, with gating times $\sim 1\mbox{ ns}$ \cite{coish:prb07}.

\subsubsection{Dynamic Nuclear Polarization and Alternative Host Materials}

For both qubit encoding schemes, control over the nuclear spin bath is of great benefit. Various methods for dynamic nuclear polarization (DNP) have been developed and demonstrated \cite{foletti:nph09, bluhm:prl10, petta:prl08, vink:nph09, danon:prl09, barthel:arX11, laird:prl07, laird:sst09}, all of which may be used to generate large magnetic field gradients between neighboring QDs. Such gradients can be measured quantitatively via the cycle duration of oscillations $\ket{S} \leftrightarrow \ket{T_0}$ \cite{foletti:nph09, bluhm:prl10, barthel:arX11}, and differences $> \mbox{200 mT}$ in the Overhauser field have been induced and subsequently exploited for coherent rotations on the Bloch sphere \cite{foletti:nph09}. In addition, schemes exist to reduce the width of the nuclear field distribution \cite{vink:nph09, danon:prl09, bluhm:prl10}, allowing to strongly prolong the dephasing times \cite{bluhm:prl10}. 

Besides GaAs QDs, other promising host materials are under investigation. In particular, rapid progress has been made on Si QDs within Si/SiGe heterostructures \cite{borselli:apl11, thalakulam:apl10, shi:arX11, simmons:prl11, prance:prl12, maune:nat12}. For instance, slow electron spin relaxation $T_1 > 1\mbox{ s}$ has recently been observed in this material via single-shot readout \cite{simmons:prl11}. Also, single-shot measurements have allowed to extract relaxation times $T_1 \sim 10\mbox{ ms}$ for $S$-$T_0$ qubits operated near the (1,1)-(0,2) charge state transition \cite{prance:prl12}. Moreover, a hyperfine-induced dephasing time $T_2^* = 360\mbox{ ns}$ has been deduced from ensemble-averaged measurements on an $S$-$T_0$ qubit \cite{maune:nat12}, which is nearly two orders of magnitude longer than the dephasing times measured in GaAs. 

As opposed to self-assembled QDs, gate-defined systems cannot confine electrons and holes at the same time and are therefore optically inactive \cite{gywat:book}. Currently, experiments are almost exclusively carried out on gated 2DEGs, and new experimental challenges may be encountered when 2D hole gases are used instead \cite{grbic:apl05}. However, considering the long lifetimes predicted for heavy-holes in QDs \cite{bulaev:prl05, fischer:prb08, fischer:prl10} and the fact that strong magnetic field gradients may be induced via micromagnets \cite{laird:prl07, laird:sst09, pioroladriere:nph08, shin:prl10, brunner:prl11}, we think that hole spins can present a valuable alternative to electron spins.

\subsection{Quantum Dots in Nanowires}

Semiconducting nanowires attracted a lot of interest as promising platforms for Majorana fermions \cite{sau:prb10, oreg:prl10, gangadharaiah:prl11}, field effect transistors \cite{xiang:nat06}, programmable circuits \cite{yan:nat11}, single-photon sources \cite{reimer:jnanophot11}, lasers \cite{yan:nphot09, chu:natnano11}, and others. QDs therein form when the confinement in the transverse directions, provided by the wire geometry, is supplemented with an additional confinement in the longitudinal direction, which can be achieved both via electric gates and via structured growth of materials with suitable band offsets. In the first case, coupling between neighboring QDs is easily controllable via the gate voltages, while the second case is highly attractive for optical processes because electrons and holes can be stored at the same time (see also self-assembled QDs). Nanowires are thus versatile and may present a valuable link between ``stationary'' and ``flying'' qubits. For instance, similarly to self-assembled QDs \cite{warburton:nat00, gerardot:book, urbaszek:arX12, gywat:book}, voltage-controlled charging and spin-dependent selection rules have been demonstrated on optically active InAsP QDs embedded in InP wires \cite{vankouwen:nanolett10, vanweert:apl10, vanweert:nanolett09}.

Nanowires have been grown from a variety of materials, all of which feature different properties and advantages. A prominent host material is InAs, which is known for its strong SOI and large $g$ factors. Experiments on electrons in InAs nanowire QDs revealed a spin-orbit length $l_{\rm SO} \simeq 130\mbox{ nm}$, i.e., a spin-orbit energy $E_{\rm SO} \simeq 100\mbox{ $\mu$eV}$ \cite{fasth:prl07}, along with $g$ factors $|g| \simeq \mbox{7-9}$ \cite{fasth:prl07, nadjperge:nat10, schroer:prl11}. These features allow for fast spin rotations via EDSR \cite{schroer:prl11, nadjperge:nat10}, and spin-flip times $<10\mbox{ ns}$ have already been demonstrated on single-electron qubits \cite{nadjperge:nat10}. On the other hand, the coherence time in Ref.\ \cite{nadjperge:nat10} was found to be rather short, only $50\mbox{ ns}$ for a single echo pulse and $<200\mbox{ ns}$ for several pulses, which is considered to be attributable to the large nuclear spin 9/2 of In. The relaxation times clearly exceeded the measurement range, i.e., the $\mu$s time scale \cite{nadjperge:nat10}, consistent with the theoretically predicted $T_1$ of several $\mu$s to ms \cite{trif:prb08}. Experiments with singlet-triplet states in an InAs nanowire DQD in the multi-electron regime showed that spin relaxation may be suppressed via tuning of the interdot coupling \cite{pfund:prl07}. Besides InAs, also InSb has recently attracted a lot of attention, where EDSR spectroscopy on the two-electron states of a gate-defined nanowire DQD revealed $l_{\rm SO} \simeq \mbox{200-300 nm}$ and very large $g$ factors $|g| \gtrsim 30$ \cite{nadjperge:arX12}.

\begin{figure}[htb]
\begin{center}
\includegraphics[width=1.00\linewidth]{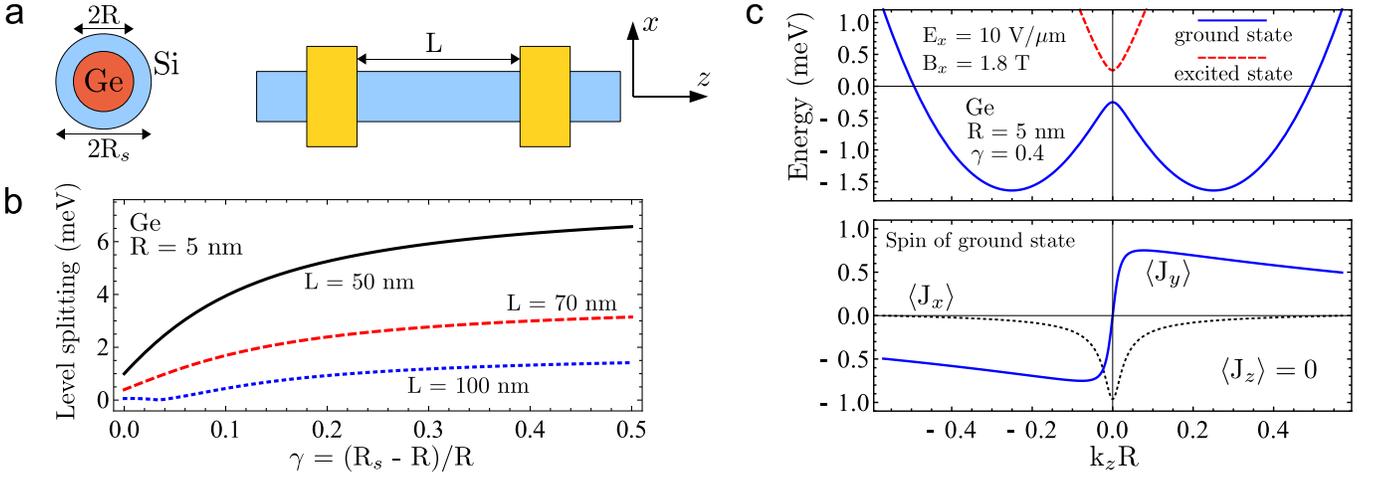}
\caption{Calculated low-energy properties of hole states in Ge/Si core/shell nanowires, following Ref.\ \cite{kloeffel:arX11}. (a) Holes accumulate in the Ge core of radius $R$, surrounded by a Si shell of thickness $R_s - R$ leading to static strain. When the nanowire is supplemented with confinement along the wire axis $z$, a quantum dot of radius $R$ and (effective) length $L$ forms. 
(b) Energy gap between the two lowest Kramers doublets in a longitudinal Ge quantum dot of $R = 5\mbox{ nm}$ and different lengths $L$, as a function of relative shell thickness $\gamma = (R_s - R)/R$. The Si shell allows to change the level splitting by several meV, which is particularly useful for implementing spin qubits. 
(c) Top: Hole spectrum as a function of longitudinal wave number $k_z$ for a typical Ge/Si wire of $R = 5\mbox{ nm}$ and $R_s = 7\mbox{ nm}$ in the presence of an electric field $E_x = 10\mbox{ V/$\mu$m}$ and a magnetic field $B_x = 1.8\mbox{ T}$ along $x$ (see panel a), illustrating strong spin-orbit interaction and sensitivity to magnetic fields, which are prerequisites for efficient qubit manipulation. The spin-orbit energy $E_{\rm SO} > 1.5 \mbox{ meV}$, resulting from direct, dipolar coupling to $E_x$, is more than 15 times larger than the reported value for InAs \cite{fasth:prl07}. At $k_z=0$, $B_x$ opens a Zeeman gap of $\mbox{0.5 meV}$, corresponding to $|g| \sim 5$. Bottom: Expectation value of the effective hole spin components $J_x$, $J_y$, and $J_z$ for the ground state plotted above. When the Fermi level is set within the gap at $k_z = 0$, the wire transports opposite spins in opposite directions, the characterisitic feature of a helical mode. Therefore, Ge/Si nanowires also provide a promising basis for spin filters and Majorana fermions \cite{sau:prb10, oreg:prl10, gangadharaiah:prl11, xiang:natnano06, kloeffel:arX11}.}
\label{figureNanowires}
\end{center}
\end{figure} 

A promising alternative to the III-V compounds are Ge/Si core/shell nanowires, which can be grown nuclear-spin-free. The valence band offset at the Ge/Si interface is large, $\sim 0.5\mbox{ eV}$, so that holes accumulate naturally in the Ge core without the need for dopants \cite{park:nanolett10, lu:pnas05}. High mobilities \cite{xiang:nat06} and very long mean free paths \cite{lu:pnas05} have been observed, along with a highly coherent charge transport seen through proximity-induced superconductivity \cite{xiang:natnano06}. Progress has also been made on gate-controlled Ge/Si nanowire QDs, even though the single-hole regime has not quite been reached yet \cite{roddaro:prl08, hu:natnano07, hu:unpubl11}. Experiments on such QDs have recently revealed spin relaxation times near 1 ms \cite{hu:unpubl11}. Furthermore, as illustrated in Fig.\ \ref{figureNanowires}b, the relative thickness of the Si shell allows to vary the QD level splitting by several meV via the static strain \cite{kloeffel:arX11}. Remarkably, Ge/Si nanowires were also found to feature strong SOI: theoretical studies of the low-energy hole states showed that direct, dipolar coupling to an electric field gives rise to a SOI of Rashba type (``direct Rashba spin-orbit interaction'') which exceeds the standard Rashba SOI by one to two orders of magnitude \cite{kloeffel:arX11}. Figure \ref{figureNanowires}c (top) plots the ground state spectrum for a typical nanowire of $\mbox{5 nm}$ core radius and $\mbox{2 nm}$ shell thickness in a moderate electric field of $10\mbox{ V/$\mu$m}$ perpendicular to the wire. The corresponding spin-orbit energy is $E_{\rm SO} > 1.5\mbox{ meV}$, more than 15 times larger than the reported value for InAs \cite{fasth:prl07}. The additional magnetic field in Fig.\ \ref{figureNanowires}c (top) along the electric field axis opens a gap in the spectrum ($|g| \simeq 5$ at $k_z = 0$), illustrating sensitivity to magnetic fields, a prerequisite for efficient qubit manipulation. The hole $g$ factors in Ge/Si nanowire QDs are tunable both via the confinement and the magnetic field orientation \cite{roddaro:prl08, kloeffel:arX11}. All these properties should allow for electrically controlled qubits with long lifetimes and short operation times. Furthermore, Ge/Si nanowires present an outstanding platform for helical hole states and Majorana fermions \cite{sau:prb10, oreg:prl10, gangadharaiah:prl11, xiang:natnano06, kloeffel:arX11}. Finally, we mention that helical states \cite{klinovaja:prb11, klinovaja:prl11} and Majorana fermions \cite{klinovaja:arX12} can alternatively be realized in armchair carbon nanotubes without the need for a magnetic field, allowing for purely electrical setups. The required field, on the order of V/nm \cite{klinovaja:prb11, klinovaja:prl11, klinovaja:arX12}, is stronger than in the case of Ge/Si wires, but still below experimentally achievable limits.

\subsection{Overview}
\label{secsub:OverviewProgress}

%%%%%%%%%%%%%%%%%%%%%%%%%%%%%%%%%%%%%%%%%%%%%%%%%%%%%%%%%%%%%%%%
%% Below: Table with references ordered by appearence in code %%
%%%%%%%%%%%%%%%%%%%%%%%%%%%%%%%%%%%%%%%%%%%%%%%%%%%%%%%%%%%%%%%%

\begin{table}[htbp] 
\caption{Overview on the state of the art for quantum computing with spins in QDs in April 2012. For each of the systems discussed in the main text, the table summarizes the longest lifetimes, the shortest operation times, the highest readout fidelilities (visibilities), and the highest initialization fidelities reported so far in experiments. Information on established schemes for readout and initialization is provided, along with a rating on scalability. All single-qubit operation times correspond to rotations of $\pi$ (about the $z$ and $x$ axis, respectively) on the Bloch sphere: for a qubit with eigenstates $\ket{0}$ and $\ket{1}$, $\tau_Z$ refers to operations of type $\left( \ket{0} + \ket{1} \right)/\sqrt{2} \to \left(\ket{0} - \ket{1}\right)/\sqrt{2}$, while $\tau_X$ refers to rotations of type $\ket{0} \to \ket{1}$. Two-qubit gates are characterized by the SWAP time $\tau_{\rm SW}$, describing operations of type $\ket{01} \to \ket{10}$. The ratio $T_2/\tau_{\rm op}$, where $\tau_{\rm op}$ is the longest of the three operation times, gives an estimate for the number of qubit gates the system can be passed through before coherence is lost. Referring to standard error correction schemes, this value should exceed $\sim 10^4$ for fault-tolerant quantum computation to be implementable. Using the surface code, values above $\sim 10^2$ may already be sufficient. Experiments on self-assembled QDs have predominantly been carried out in (In)GaAs. Unless stated otherwise, GaAs has been the host material for gate-defined QDs in 2DEGs. The results listed for nanowire QDs have been achieved in InAs (electrons) and Ge/Si core/shell (holes) nanowires. We note that the experimental conditions, such as externally applied magnetic fields, clearly differ for some of the listed values and schemes. Entries ``n.a.'' stand for ``not yet available''. Finally, we wish to emphasize that further improvements might already have been achieved of which we had not been aware when writing this review.}
\vspace{0.15cm}
\begin{tabular}{>{\raggedright}m{0.109\linewidth} | >{\centering}m{0.138\linewidth} | >{\centering}m{0.138\linewidth} | >{\centering}m{0.138\linewidth} | >{\centering}m{0.138\linewidth} | >{\centering}m{0.138\linewidth} | >{\centering}m{0.138\linewidth} |} 

\multicolumn{1}{c|}{} &\multicolumn{2}{c|}{\bf Self-Assembled QDs}& \multicolumn{2}{c|}{\bf Lateral QDs in 2DEGs}&\multicolumn{2}{c|}{\bf QDs in Nanowires }  \tn  

\multicolumn{1}{c|}{} & Electrons & Holes & Single Spins & $S$-$T_0$ Qubits & Electrons & Holes \tn  \hline \hline

Lifetimes &
\mbox{ }\vspace{-0.26cm}\\
$\begin{array}{l} T_1 > 20\mbox{ ms\tss{a}} \\ T_2: \mbox{ 3 $\mu$s\tss{b}} \\  T_2^* \gtrsim \mbox{ 0.1 $\mu$s\tss{c}} \\ \mbox{ } \\ \mbox{ }  \end{array}$ & 
\mbox{ }\vspace{-0.26cm}\\  
$\begin{array}{l} T_1: \mbox{ 0.5 ms\tss{d}} \\ T_2: \mbox{ 1.1 $\mu$s\tss{e}} \\ T_2^* > 0.1\mbox{ $\mu$s\tss{f}} \\ \mbox{ } \\ \mbox{ } \end{array}$  & 
\mbox{ }\vspace{-0.26cm}\\
$\begin{array}{l} T_1 > 1\mbox{ s\tss{g}} \\ T_2: \mbox{ 0.44 $\mu$s\tss{h}} \\ T_2^*: \mbox{ 37 ns\tss{h}} \\ T_1^{\rm Si} > 1\mbox{ s\tss{i}} \\ \mbox{ } \end{array}$  &
\mbox{ }\vspace{-0.26cm}\\
$\begin{array}{l} T_1: \mbox{ 5 ms\tss{j}} \\ T_2: \mbox{ 276 $\mu$s\tss{k}} \\ T_2^*: \mbox{ 94 ns\tss{l}} \\ T_1^{\rm Si} \sim \mbox{ 10 ms\tss{m}} \\ T_2^{*, {\rm Si}}: \mbox{ 360 ns\tss{n}}  \end{array}$   &
\mbox{ }\vspace{-0.26cm}\\
$\begin{array}{l} T_1 \gg 1\mbox{ $\mu$s\tss{o}} \\ T_2: \mbox{ 0.16 $\mu$s\tss{o}} \\ T_2^*: \mbox{ 8 ns\tss{o}} \\ \mbox{ }  \\ \mbox{ } \end{array}$ &
\mbox{ }\vspace{-0.26cm}\\
$\begin{array}{l} T_1: \mbox{ 0.6 ms\tss{p}} \\ T_2: \mbox{ n.a.} \\ T_2^*: \mbox{ n.a.} \\ \mbox{ } \\ \mbox{ } \end{array}$  \tn \hline

Operation times & 
\mbox{ }\vspace{-0.26cm}\\
$\begin{array}{l} \tau_Z: \mbox{ 8.1 ps\tss{q}} \\ \tau_X:\mbox{ 4 ps\tss{r}}\\ \tau_{\rm SW}: \mbox{17 ps\tss{s}} \end{array}$ & 
\mbox{ }\vspace{-0.26cm}\\
$\begin{array}{l} \tau_Z: \mbox{ 17 ps\tss{e}} \\ \tau_X:\mbox{ 4 ps\tss{e}}\\ \tau_{\rm SW}: \mbox{25 ps\tss{t}} \end{array}$  & 
\mbox{ }\vspace{-0.26cm}\\
$\begin{array}{l} \tau_Z: \mbox{ n.a.\tss{u}} \\ \tau_X:\mbox{ 20 ns\tss{v}}\\ \tau_{\rm SW}: \mbox{ 350 ps\tss{w}} \end{array}$  & 
\mbox{ }\vspace{-0.26cm}\\
$\begin{array}{l} \tau_Z: \mbox{ 350 ps\tss{w}} \\ \tau_X:\mbox{ 0.39 ns\tss{x}}\\ \tau_{\rm ccpf}: \mbox{ 30 ns\tss{y}} \end{array}$  & 
\mbox{ }\vspace{-0.26cm}\\
$\begin{array}{l} \tau_Z: \mbox{ n.a.\tss{z}} \\ \tau_X:\mbox{ 8.6 ns\tss{o}}\\ \tau_{\rm SW}: \mbox{ n.a.} \end{array}$  & 
n.a. \tn \hline

\mbox{ }\vspace{-0.32cm}\\ $T_2/\tau_{\rm op}$ & \mbox{ }\vspace{-0.32cm}\\ $1.8\times 10^5$  & \mbox{ }\vspace{-0.32cm}\\ $4.4\times 10^4$ & \mbox{ }\vspace{-0.32cm}\\ 22 & \mbox{ }\vspace{-0.32cm}\\ $9.2\times 10^3$ & \mbox{ }\vspace{-0.32cm}\\ n.a. & \mbox{ }\vspace{-0.32cm}\\ n.a. \tn \hline

Readout \& fidelities $F$ (visibilities $V$) & 
\mbox{ }\vspace{-0.25cm}\\
$F = 96\%$\tss{A} \\ Resonance fluorescence in a QD molecule \\ \mbox{ }\\ Other:\\ Faraday\tss{B} and Kerr\tss{C} rotation spectroscopy, resonance fluorescence\tss{D} &  
Absorption\tss{E} and emission\tss{F} spectroscopy\\ (selection rules)  & 
\mbox{ }\vspace{-0.30cm}\\
$V = 65\%$\tss{G}\\$V^{\rm Si} = 88\%$\tss{i} \\ Spin-selective tunneling \\ \mbox{ }\\ $F = 86\%$\tss{H}\\ Spin-selective tunneling \\(two spins) \\ \mbox{ }\\ Other: \\Photon-assisted tunneling\tss{I}  & 
\mbox{ }\vspace{-0.30cm}\\ 
$V = 90\%$\tss{J} \\ $F = 97\%$\tss{K} \\ Spin-dependent charge distribution (rf-QPC\tss{J}/SQD\tss{K}) \\ \mbox{ }\\ $V = 81\%$\tss{L} \\ Spin-dependent tunneling rates \\ \mbox{ }\\ Other: \\Dispersive readout\tss{M}  &
$F = \mbox{70-80\%}$\tss{N} Pauli spin blockade  & Spin-dependent charge distribution (sensor dot coupled via floating gate)\tss{O}  \tn \hline

Initialization \& fidelities $F_{\rm in}$ & 
$F_{\rm in} > 99\%$\tss{P} \\ Optical pumping  & 
$F_{\rm in} = 99\%$\tss{Q} \\ Optical pumping \\ $F_{\rm in} > 99\%$\tss{R}\\ Exciton ionization &  
\mbox{ }\vspace{-0.33cm}\\ Spin-selective tunneling,\tss{S} adiabatic ramping to ground state of nuclear field\tss{T} & 
Pauli exclusion\tss{w} & 
Pauli spin blockade\tss{N} & 
n.a. \\(single-hole regime not yet reached)  \tn \hline

Scalability &\multicolumn{2}{c|}{Scaling seems challenging}& \multicolumn{2}{c|}{$\begin{array}{c} \mbox{ }\vspace{-0.32cm}\\ \mbox{Seems scalable\tss{U}} \\ \mbox{(e.g.\ via floating gates\tss{V})} \end{array}$ } & \multicolumn{2}{c|}{$\begin{array}{c} \mbox{ }\vspace{-0.32cm}\\ \mbox{Seems scalable} \\ \mbox{(e.g.\ via floating gates\tss{V})} \end{array}$}  \tn \hline

\end{tabular}

\vspace{-0.25cm}
\begin{tabular}{>{\raggedright}m{0.98\linewidth}}
\mbox{ } \\
{\footnotesize \tss{a}\cite{kroutvar:nat04}; \tss{b}\cite{greilich:sci06, press:nphot10}; \tss{c}Measured for single electrons in a narrowed nuclear spin bath \cite{xu:nat09} and for two-electron states in QD molecules with reduced sensitivity to electrical noise \cite{weiss:arX12}, both via coherent population trapping. Without preparation $T_2^* \sim \mbox{0.5-10 ns}$, see Sec.\ \ref{secsub:ProgressSelfAssembled}.; \tss{d}\cite{gerardot:nat08, heiss:prb07}; \tss{e}\cite{degreve:nph11}; \tss{f}\cite{brunner:sci09}. Measured through coherent population trapping. Other experiments revealed $T_2^* = \mbox{ 2-21 ns}$ attributed to electrical noise \cite{degreve:nph11, greilich:arX11}.; \tss{g}\cite{amasha:prl08};  \tss{h}\cite{koppens:prl08}; \tss{i}\cite{simmons:prl11}; \tss{j}\cite{johnson:nat05, hanson:prl05}; \tss{k}\cite{bluhm:nph11}; \tss{l}\cite{bluhm:prl10}. Achieved by narrowing the nuclear spin bath, $T_2^* \sim\mbox{10 ns}$ otherwise \cite{bluhm:prl10, petta:sci05, johnson:nat05}.; \tss{m}\cite{prance:prl12}; \tss{n}\cite{maune:nat12}. Hyperfine-induced.; \tss{o}\cite{nadjperge:nat10}; \tss{p}\cite{hu:unpubl11}; \tss{q}\cite{press:nphot10}; \tss{r}\cite{press:nat08,press:nphot10}; \tss{s}\cite{kim:nph11}; \tss{t}\cite{greilich:arX11}; \tss{u}While Ref.\ \cite{koppens:prl08} gets close, we are currently not aware of a Ramsey-type experiment where coherent rotations about the Bloch sphere $z$ axis have explicitly been demonstrated as a function of time. Thus no value is listed. However, $\tau_Z$ should be short, on the order of 0.1 ns assuming a magnetic field of 1 T and $g = -0.44$ as in bulk GaAs.; \tss{v}\cite{pioroladriere:nph08}; \tss{w}\cite{petta:sci05}; \tss{x}\cite{foletti:nph09}; \tss{y}\cite{vanweperen:prl11}. SWAP gates for $S$-$T_0$ qubits have not yet been implemented so far. We therefore list the duration $\tau_{\rm ccpf}$ of a charge-state conditional phase flip.; \tss{z}In Ref.\ \cite{nadjperge:nat10}, rotations about an arbitrary axis in the $x$-$y$ plane of the Bloch sphere are reported instead, controlled via the phase of the applied microwave pulse.; \tss{A}\cite{vamivakas:nat10}; \tss{B}\cite{atatuere:nph07}; \tss{C}\cite{berezovsky:sci08, mikkelsen:nph07, berezovsky:sci06}; \tss{D}\cite{vamivakas:nph09}; \tss{E}\cite{gerardot:nat08, greilich:arX11}; \tss{F}\cite{heiss:prb07, degreve:nph11}; \tss{G}\cite{elzerman:nat04}; \tss{H}\cite{nowack:sci11}; \tss{I}\cite{shin:prl10}; \tss{J}\cite{barthel:prl09}. rf-QPC: radio-frequency quantum point contact \cite{reilly:apl07, cassidy:apl07}.; \tss{K}\cite{shulman:arX12}. rf-SQD: radio-frequency sensor quantum dot \cite{barthel:prb10}.; \tss{L}\cite{hanson:prl05}. The paper demonstrates readout of the singlet and triplet states in a single quantum dot.; \tss{M}\cite{petersson:nanolett10}. A radio-frequency resonant circuit is coupled to a double quantum dot.; \tss{N}\cite{nadjperge:nat10}. The readout and initialization schemes in this experiment only determine whether two neighboring spin qubits are equally or oppositely oriented.; \tss{O}\cite{hu:unpubl11}. The scheme, operated in the multi-hole regime, distinguishes the spin triplet states from the spin singlet.; \tss{P}\cite{atatuere:sci06}; \tss{Q}\cite{gerardot:nat08}; \tss{R}\cite{godden:apl10}; \tss{S}\cite{elzerman:nat04, nowack:sci11, simmons:prl11}; \tss{T}\cite{petta:sci05}. Information about the nuclear field is required for the ground state to be known.; \tss{U}\cite{gaudreau:nph12}; \tss{V}\cite{trifunovic:arX11}.} 
\end{tabular}

\label{TableComparison}
\end{table}

Table \ref{TableComparison} summarizes important information about the systems covered in this section, such as the longest measured lifetimes and shortest reported operation times (as of April 2012). We would like to point out that, even though we studied the literature carefully, the provided summary is not intended to be complete, as further improvements might already have been achieved of which we had not been aware when completing this review. The table also lists established initialization and readout schemes for each system, along with a ratio of the observed decoherence and gating times. The latter illustrates that decoherence no longer presents the massive stumbling block it had been considered years ago.

Typical values for characterizing readout schemes are the measurement fidelities $F$ and the visibilities $V$. We note that these are not exactly equivalent, as briefly explained in the following. When $e_{0 \to 1}$ denotes the error probability that the qubit state $\ket{0}$ is incorrectly read as $\ket{1}$, the measurement fidelity for state $\ket{0}$ is $F_0 = 1 - e_{0 \to 1}$ \cite{elzerman:nat04, hanson:prl05, barthel:prl09}. Analogously, the measurement fidelity for the qubit state $\ket{1}$ is $F_1 = 1 - e_{1 \to 0}$. One may define the readout fidelity for a particular experiment as $F = p_0 F_0 + p_1 F_1 = 1 - p_0 e_{0 \to 1} - p_1 e_{1 \to 0}$, where $p_0$ ($p_1$) is the probability that the system is initially in $\ket{0}$ ($\ket{1}$) \cite{vamivakas:nat10}. Weighting $\ket{0}$ and $\ket{1}$ equally, this results in $F = 1 - (e_{0 \to 1} + e_{1 \to 0})/2$ \cite{hanson:rmp07, meunier:prb06}. A more general quantity is the visibility $V = 1 - e_{0 \to 1} - e_{1 \to 0}$, which is independent of $p_0$ and $p_1$ and presents a lower bound for the readout fidelity in a system \cite{elzerman:nat04, hanson:prl05, barthel:prl09, hanson:rmp07}. We note that readout may also be characterized by the measurement efficiency defined in Ref.\ \cite{engel:prl04}, where various readout schemes have been analyzed theoretically.

\section{Proposals for Long-Distance Spin-Spin Coupling}
\label{sec:NewProposals}

\begin{figure}[htb]
\begin{center}
\includegraphics[width=1.00\linewidth]{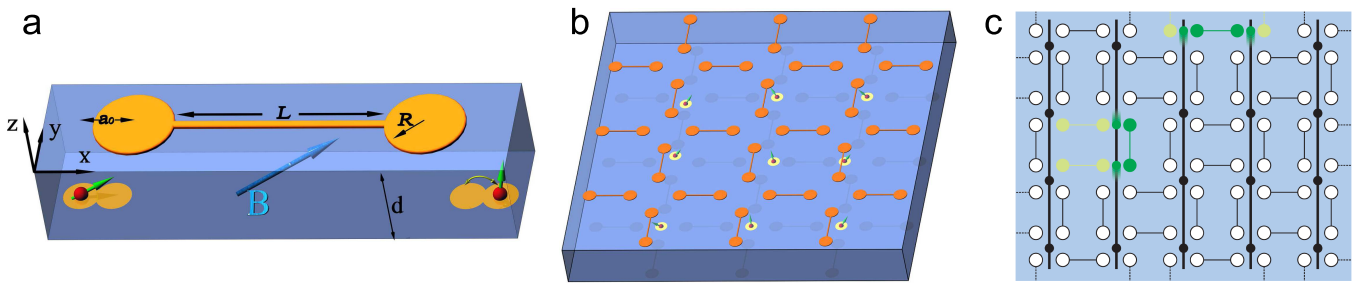}
\caption{Long-distance qubit-qubit coupling via floating gates, allowing to place individual qubits far apart \cite{trifunovic:arX11}. (a) Schematic setup for two spins in separated  double quantum dots coupled capacitively via a floating gate (here: simple, symmetric ``dog bone'' geometry). In the presence of a magnetic field and spin-orbit coupling, the electrostatic interaction of the charges results in an effective spin-spin coupling. Depending on the actual system and the gate geometry, remarkably strong qubit-qubit interactions of $\mbox{1-100 $\mu$eV}$ can be reached \cite{trifunovic:arX11}. \mbox{(b) Scalable} quantum computer architecture using metallic floating gates on top of a two-dimensional electron gas. (c) An alternative architecture with qubits (black dots) implemented in nanowires (vertical black lines). In both (b) and (c), qubit-qubit interactions can be switched on (off) via gates by moving the qubits close to (away from) the corresponding metal discs. The architectures provide a platform for the powerful surface code \cite{raussendorf:njp07, raussendorf:prl07, fowler:pra09, wang:pra11, divincenzo:physscr09, wootton:arX12, yao:nat12}. {\it All pictures were taken from \cite{trifunovic:arX11} and are property of the American Physical Society (aps.org).} }
\label{figureFloatingGates}
\end{center}
\end{figure}

Recently, pairwise control of the exchange interaction via electric gates has been demonstrated in a triple quantum dot \cite{gaudreau:nph12, koppens:nph12}. Such control is an essential requirement for most quantum computer architectures (see Fig.\ \ref{schemeLDiVQC}b), and so the experiments of Ref.\ \cite{gaudreau:nph12} present an important proof of scalability. Large-scale quantum computers, however, must be capable of reaching a system size of several thousands of qubits. This poses serious architectural challenges to the exchange-based QD scheme from Ref.\ \cite{loss:pra98}, Fig.\ \ref{schemeLDiVQC}, since the large amount of wires and metallic gates needs to be installed and operated on a very small scale. A promising strategy to meet this challenge has recently been proposed \cite{trifunovic:arX11}; long-distance spin-spin coupling can be achieved capacitively via floating gates, allowing to move the (D)QDs far apart. The effective qubit-qubit coupling $J^\prime$ via floating gates can take remarkably large values $J^\prime \sim \mbox{1-100 $\mu$eV}$ \cite{trifunovic:arX11}. These are comparable to the achievable exchange energies $J \sim \mbox{10 - 100 $\mu$eV}$ in typical GaAs DQDs \cite{hanson:rmp07, loss:pra98, trifunovic:arX11}, where we note that $J$ close to \mbox{10 $\mu$eV} has already been realized \cite{petta:sci05}. The floating gates may be positioned on top of the sample, as sketched in Fig.\ \ref{figureFloatingGates}a, or may even be defined within the 2DEG. Qubit-qubit coupling can be switched on and off by changing the relative positions of the QDs (charges) with respect to the gates, allowing for all-electrical control \cite{trifunovic:arX11}. Proposed, scalable architectures for a quantum computer with floating gates are shown in Figs.\ \ref{figureFloatingGates}b-c. A key feature of the architectures suggested in Ref.\ \cite{trifunovic:arX11} is that they all consist of a two-dimensional lattice of spin qubits with nearest neighbor qubit-qubit interactions. Therefore, they all allow for the implementation of the surface code with its strikingly large error threshold around 1\% (see also Sec.\ \ref{sec:RecentProgress} and Refs.\ \cite{raussendorf:njp07, raussendorf:prl07, fowler:pra09, wang:pra11, divincenzo:physscr09, wootton:arX12, yao:nat12} for further information).  

\begin{figure}[htb]
\begin{center}
\includegraphics[width=1.00\linewidth]{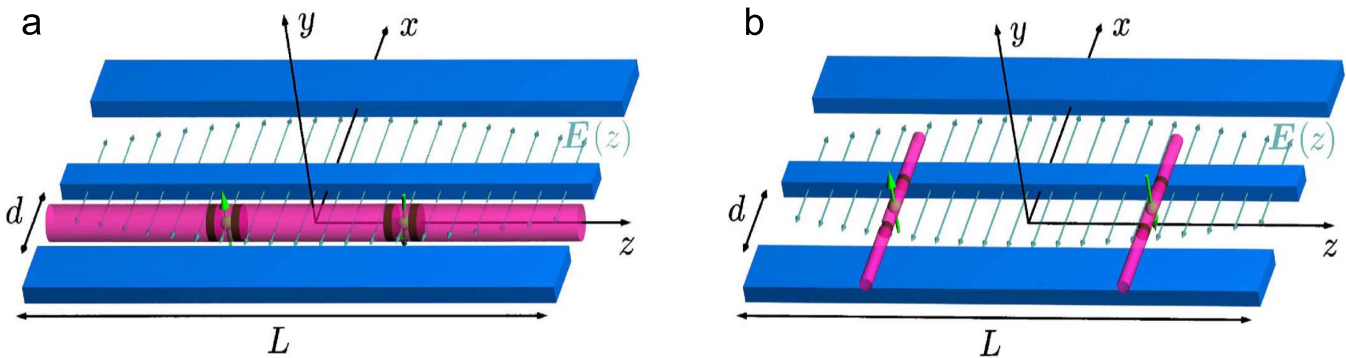}
\caption{Architectures for long-range qubit-qubit coupling via microwave cavities as studied in Ref.\ \cite{trif:prb08}. The superconducting transmission line resonator is sketched in blue, assuming that the center conductor of length $L$ is separated from the neighboring ground planes by a \mbox{distance $d$}. Electron spin states (thick green arrows) in InAs nanowire quantum dots (QDs) serve as qubits. The QD confinement, illustrated by brown discs within the nanowires (pink cylinders), can be realized with a suitable barrier material, such as InP, or with electric gates. (a) A large-diameter nanowire is oriented parallel to the transmission line and hosts QDs with strong longitudinal confinement. (b) Small-diameter nanowires, each hosting a QD with strong transverse confinement, are placed perpendicular to the transmission line. An insulating material separates the wires from the superconducting resonator to prevent a current flow. In both setups, the spin-orbit interaction enables all-electrical operation along with strong long-range interactions mediated by the photon field of the cavity \cite{trif:prb08}. For the system in (b), the time to coherently swap two spins was estimated to be around \mbox{20 ns}, and further improvement seems clearly possible. {\it The pictures were taken from \cite{trif:prb08} and are property of the American Physical Society (aps.org).} }
\label{figureStriplinesInAs}
\end{center}
\end{figure}  

Another scheme for long-distance coupling between spins uses the photon field of a cavity. The original idea goes back to a proposal from 1999, which suggests that laser-induced Raman transitions can be used to couple the electron spin states of distant QDs when embedded in an optical cavity \cite{imamoglu:prl99}. A few years later, as an alternative to the standard cavity quantum electrodynamics (QED), 1D superconducting transmission line resonators, which operate as on-chip microwave cavities, have been introduced and have launched the field of circuit QED \cite{blais:pra04, wallraff:nat04}. Since then, several proposals for long-range spin-spin coupling via circuit QED have been made \cite{childress:pra04, burkard:prb06, taylor:arX06, trif:prb08, cottet:prl10, hu:arX12}. The direct coupling of a single spin to the magnetic component of the cavity electromagnetic field, however, is weak, and achieving strong interactions between a spin qubit and a cavity thus requires other mechanisms, resulting in a variety of suggested approaches. These, for instance, are based on Raman-type transitions among single-electron states in DQDs \cite{childress:pra04}, on $S$-$T_0$ qubits in QD molecules with (nuclear) magnetic field gradients \cite{burkard:prb06, taylor:arX06}, or on the single-electron states in DQDs contacted to ferromagnetic insulators \cite{cottet:prl10}. Relatively recently, it has been proposed to use electron spins in InAs nanowire QDs, which feature a strong SOI and thus enable efficient coupling between the qubits and the electric component of the cavity field \cite{trif:prb08}. Investigated architectures are shown in Fig.\ \ref{figureStriplinesInAs}, where nanowires with strong longitudinal (transverse) QD confinement are placed parallel (perpendicular) to the transmission line in panel a (b). Rotations of individual qubits can be driven through EDSR \cite{golovach:prb06, nadjperge:nat10, nadjperge:arX12, schroer:prl11}, and two-qubit interactions can be turned on and off by changing the QD confinements with nearby gates, allowing for all-electrical control \cite{trif:prb08}. For the setup of Fig.\ \ref{figureStriplinesInAs}b, an operation time around \mbox{20 ns} was estimated for swapping two spins coherently, and further optimization seems clearly possible. Notably, it has been shown that circuit QED also allows for long-range coupling between certain types of molecular magnets \cite{trif:prl08}.

\section{Outlook}
\label{sec:Outlook}

When electron spins in QDs were proposed for quantum computation, the experimental situation was not encouraging. Gate-controlled QDs within 2DEGs were limited to around 30 or more confined electrons each, and techniques for single-qubit manipulation and readout were not available \cite{sachrajda:vpoint11}. Furthermore, decoherence from interactions with the environment was considered an almost insurmountable obstacle. Within the past decade, the situation has changed dramatically, owing to continuous experimental and theoretical progress. QDs are now routinely controlled down to the last electron (hole), and various schemes have been applied both for qubit initialization and readout. Reducing the occupation number of QDs to the minimum is desirable for high-fidelity quantum computation \cite{hu:pra01}; however, larger fillings with a well-defined spin-1/2 ground state are also useful \cite{meier:prl03}. In addition, efficient single- and two-qubit gates have been demonstrated, allowing for universal quantum computing when combined. The achieved gating times are much shorter than measured lifetimes, and it seems that one will soon be able to overcome decoherence to the required extent. This is a major step toward the realization of a quantum computer. 

While the field is very advanced for the ``workhorse'' systems such as lateral GaAs QDs or self-assembled (In)GaAs QDs, rapid progress is also being made in the quest for alternative systems with further optimized performance. First, this includes switching to different host materials. For instance, Ge and Si can be grown nuclear-spin-free, and required gradients in the Zeeman field may be induced via micromagnets. Second, both electron- and hole-spin qubits are under investigation, exploiting the different properties of conduction and valence band, respectively. Finally, promising results are obtained from new system geometries, particularly nanowire QDs.

Future tasks can probably be divided into three categories. The first one consists of studying new quantum computing protocols (such as the surface code) which put very low requirements on the physical qubits. The second category refers to further optimization of the individual components listed in Table \ref{TableComparison}. For instance, longer lifetimes are certainly desired, as well as high-quality qubit gates with even shorter operation times. However, as decoherence no longer seems to present the limiting issue, particular focus should also be put on implementing schemes for highly reliable, fast, and scalable qubit readout (initialization) in each of the systems. Finally, since the results in Table \ref{TableComparison} are usually based on different experimental conditions, the third category consists of merging all required elements into one scalable device, without the need for excellent performance. Such a ``complete spin-qubit processor'' should combine individual single-qubit rotations about arbitrary axes, a controlled (entangling) two-qubit operation, initialization into a precisely known state, and single-shot readout of each qubit. While Ref.\ \cite{brunner:prl11} presents an important step toward this unit, prototypes of a complete spin-qubit processor could present the basis for continuous optimization.

In summary, considering the impressive progress achieved within the past decade, one may be cautiously optimistic that a large-scale quantum computer can indeed be realized. \\ \\ \\

{\bf Summary Points:}

\begin{itemize}
\item The experimental situation has dramatically changed since 1998, when the original proposal for quantum computation with quantum dots \cite{loss:pra98} was published. Quantum dots are now precisely controlled down to the last spin. Single-qubit rotations around different axes, two-qubit operations, and various initialization and readout schemes have successfully been demonstrated (Table \ref{TableComparison}).
\item It seems that one can soon overcome decoherence to the required extent, which is a big step toward the implementation of a quantum computer.
\item Nuclear spins and spin-orbit interaction, on the one hand, present a source of decoherence and relaxation. On the other hand, they can generate large Overhauser fields and are useful for realizing quantum gates. Schemes exist to narrow the width of the nuclear field distribution. 
\item A variety of quantum dot systems is currently under investigation, including host materials with or without nuclear spins, operation in the conduction or valence band, and different geometries. Holes in Ge/Si-nanowire-based quantum dots are promising examples \cite{hu:natnano07, roddaro:prl08, hu:unpubl11, kloeffel:arX11}.
\item Electrically pulsed, pairwise control of the exchange interaction has been demonstrated in a triple quantum dot \cite{gaudreau:nph12}, i.e., the scalability of exchange-based schemes for quantum computing \cite{loss:pra98} has now been proven experimentally.  
\item Long-distance spin-spin coupling via floating gates may be used to overcome architectural challenges of a large-scale quantum computer, and several two-dimensional architectures have been proposed \cite{trifunovic:arX11}. Alternatively, distant spins may be coupled via the photon field of a cavity \cite{imamoglu:prl99, childress:pra04, burkard:prb06, taylor:arX06, cottet:prl10, trif:prb08, hu:arX12, trif:prl08}.      
\item Recent development toward quantum computation with quantum dots has been very positive, and one can be curious about the progress of the next few years. \\
\end{itemize}

{\bf Future Issues:}

\begin{itemize}
\item The surface code is a powerful protocol for fault-tolerant quantum computing \cite{raussendorf:njp07, raussendorf:prl07, wang:pra11, fowler:pra09, divincenzo:physscr09, wootton:arX12, yao:nat12}. Can the requirements on the physical qubits be reduced further? 
\item Can longer lifetimes and more efficient qubit gates be reported? In particular, can schemes for fast and highly reliable single-shot readout and initialization be implemented in each of the discussed systems? 
\item Schemes for exchange-controlled single-spin rotations (about arbitrary axes) have been proposed as an efficient alternative to rotations driven by oscillating fields \cite{loss:pra98, coish:prb07}. Can this be realized experimentally?
\item Electrically controlled single-qubit and two-qubit operations were demonstrated in the setup of Ref.\ \cite{brunner:prl11}. Can a ``complete spin-qubit processor'' be reported soon?  \\
\end{itemize}

\section*{Disclosure statement}
The authors are not aware of any affiliations, memberships, funding, or financial holdings that might be perceived as affecting the objectivity of this review.

\section*{Acknowledgments}
We thank S.\ Gangadharaiah,  F.\ Maier, P. Stano, D.\ Stepanenko, M.\ Trif, R.\ J.\ Warburton, and J.\ R.\ Wootton for helpful discussions and acknowledge support from the Swiss NF, NCCRs Nanoscience and QSIT, DARPA, and IARPA.

%\vspace{0.00cm}

\section*{List of Acronyms}

\begin{tabular}{lcl}
2DEG & : & two-dimensional electron gas \\
2DHG & : & two-dimensional hole gas \\
DNP & : & dynamic nuclear polarization \\
DQD & : & double quantum dot \\
EDSR & : & electric-dipole-induced spin resonance \\
ESR & : & electron spin resonance \\
LH & : & light-hole \\
HH & : & heavy-hole \\
QD & : & quantum dot \\
QED & : & quantum electrodynamics \\
QPC & : & quantum point contact \\
rf & : & radio-frequency \\
SOI & : & spin-orbit interaction 
\end{tabular}

\end{document}